%% file: main.tex
\documentclass[a4paper]{article}

\pdfoutput=1

\usepackage[utf8]{inputenc}
\usepackage[T1]{fontenc}
\usepackage[english]{babel}
\usepackage{csquotes}
\usepackage{pifont} 
\usepackage{amsmath, amsfonts, bm}

\usepackage[a4paper,top=3cm,bottom=2cm,left=3cm,right=3cm,marginparwidth=2cm]{geometry}
\usepackage{parskip}
\usepackage{booktabs}
\usepackage{tabularx}
\usepackage{multirow}
\usepackage{makecell}
\usepackage{graphicx}
\usepackage{subcaption}
\usepackage{threeparttable}
\usepackage{float}
\usepackage{lscape}

\usepackage[version=4]{mhchem}
\usepackage{siunitx}
\DeclareSIUnit\angstrom{\text{Å}}

\usepackage{xcolor}
\usepackage{textcomp}
\usepackage[normalem]{ulem}
\usepackage{outlines}

\usepackage{authblk}

\usepackage[backend=biber, natbib=true, style=numeric, sorting=none]{biblatex}
\usepackage{nomencl}
\usepackage[final]{changes}
\usepackage{hyperref}
\hypersetup{
    colorlinks=true,
    allcolors=blue
}

\date{}

\title{Accelerating dynamic polarizability calculations of organic molecules using equivariant graph neural networks}

\author[1,2,$\dagger$]{Houssam Metni}
\author[2,$\dagger$]{Maria Kraus}
\author[1]{Marie Louise Schubert}
\author[3]{Marjan Krsti\'{c}}
\author[1,3,$\star$]{Carsten Rockstuhl}
\author[1,2,$\star$]{Pascal Friederich}

\affil[1]{Institute of Nanotechnology, Karlsruhe Institute of Technology (KIT), Karlsruhe, Germany}
\affil[2]{Institute of Anthropomatics and Robotics, Karlsruhe Institute of Technology (KIT), Karlsruhe, Germany}
\affil[3]{Institute of Theoretical Solid State Physics, Karlsruhe Institute of Technology (KIT), Karlsruhe, Germany}

\affil[$\dagger$]{These authors contributed equally; houssam.metni@kit.edu and maria.kraus@student.kit.edu}
\affil[$\star$]{These authors are corresponding authors; carsten.rockstuhl@kit.edu and pascal.friederich@kit.edu}

\begin{document}
\maketitle

\begin{abstract}
Predicting the dynamic polarizability tensor of organic molecules is essential for simulating light–matter interactions in optoelectronic devices, yet conventional quantum-chemical methods such as time-dependent density functional theory (TD-DFT) are computationally expensive and limit large-scale screening.  
We present an equivariant graph neural network architecture that predicts frequency-dependent, complex-valued dynamic polarizability tensors directly from readily available molecular information, such as 3D geometry and UV–vis spectra.  
The model learns to approximate the shape and magnitude of the dynamic polarizability tensor, capturing both dispersive and absorptive features that define the molecular optical response.  
We evaluate the model on molecules from both the QM9 dataset and the Harvard organic photovoltaic dataset using various metrics, including the earth mover's distance, which is suitable for comparing spectral quantities.
For both datasets, the model successfully reproduces the main features of the dynamic polarizability tensor and generalizes across diverse molecules.
To assess the applicability, we further validated the model using a downstream workflow to calculate the optical properties of organic photovoltaic devices. We used the model-predicted polarizabilities to perform device-level optical simulations, yielding charge carrier generation rates in agreement with TD-DFT-based results ($R^2 = 0.94$).
Our approach offers a promising complement, or even an alternative, to conventional methods such as TD-DFT, enabling faster and more efficient screening of photoactive organic molecules for optoelectronic applications.
\end{abstract}

\begin{refsection}
\section{Introduction}
\input{1-introduction.tex}
\section{Methods}
\input{2-methods.tex}
\section{Results and discussion}
\label{Results}
\input{3-results_discussion.tex}
\section{Conclusion and outlook}
\input{4-conclusion_outlook.tex}

\section*{Data availability}
The datasets and trained model weights used in this work are available on Zenodo: \url{https://doi.org/10.5281/zenodo.23180873}. The quantum chemistry calculations are deposited at the following NOMAD repository: \url{https://doi.org/10.17172/NOMAD.RG5M-AQT6}.

\section*{Code availability}
The source code used in this work is available on the following GitHub repository: \url{https://github.com/aimat-lab/DynPolDetanet}.

\section*{Author contributions}
HM, MK, CR and PF contributed to the conceptualization. Formal analysis was performed by HM, MK, MLS and MKr. All authors contributed to the methodology, which was led by HM and MK. Investigation was performed by HM, MK, MLS and MKr. MKr performed the TD-DFT, UV–vis, T-matrix and polarizability calculations, and MLS performed the generation-rate calculations. Software was developed by HM and MK. Visualization was performed by HM and MK. HM, MK, MLS and MKr wrote the original draft. CR and PF acquired the funding and reviewed and edited the manuscript. All authors reviewed and approved the final manuscript.

\section*{Competing interests}
All authors declare no financial or non-financial competing interests. 

\section*{Acknowledgment} \par 
H.M., M.L.S., C.R., and P.F. acknowledge financial support from the Helmholtz Project 'SolarTAP'.
M.K., P.F., and C.R. acknowledge support by the Deutsche Forschungsgemeinschaft (DFG, German Research Foundation) under Germany’s Excellence Strategy via the Excellence Cluster 3D Matter Made to Order (EXC-2082/2-390761711) and from the Carl Zeiss Foundation via the CZF-Focus@HEiKA Program. M.K. and C.R. acknowledge funding by the Volkswagen Foundation. M.L.S. acknowledges the Karlsruhe School of Optics and Photonics (KSOP). 
M.K. and C.R. acknowledge support by the state of Baden--Württemberg through bwHPC and the German Research Foundation (DFG) through grant no. INST 40/575-1 FUGG (JUSTUS 2 cluster). The authors gratefully acknowledge the computing time provided on the high-performance computer HoreKa by the National High-Performance Computing Center at KIT (NHR@KIT). This center is jointly supported by the Federal Ministry of Education and Research and the Ministry of Science, Research and the Arts of Baden-Württemberg, as part of the National High-Performance Computing (NHR) joint funding program (\url{https://www.nhr-verein.de/en/our-partners}). HoreKa is partly funded by the German Research Foundation (DFG). M.K. thanks Jan David Fischbach for the help with the AAA algorithm used to interpolate the calculated dynamic polarizability tensors. 

\printbibliography
\end{refsection}

\begin{refsection}
\input{si.tex}
\end{refsection}

\end{document}

%% file: 1-introduction.tex
Organic optoelectronic materials, \textit{e.g.}, materials used for organic photovoltaics (OPV) are a promising class of materials due to their mechanical flexibility and lightweight fabrication~\cite{Zhang2025_OrganicPhotodetectors, green2014emergence, Wang2024_FlexibleOrganicOptoelectronics, Nowsherwan2024_OrganicOptoelectronics, Ostroverkhova2016_OrganicOptoelectronics, ColsmanetalOPV, friederich2017rational, wu2024inverse}. Their performance in optoelectronic devices is fundamentally determined by how they interact with light.

Optimizing the optical properties of organic semiconductors is therefore central to advancing organic optoelectronic technologies. However, this is challenging and computationally demanding. For each candidate molecule, one must determine how it interacts with electromagnetic fields across relevant excitation energies, a property encoded in the dynamic polarizability tensor. This frequency-dependent tensor quantifies how the molecular electron cloud deforms in response to incident light and thus serves as the fundamental link between molecular structure and optical behavior.

\begin{figure}[!htb]
    \centering
    \includegraphics[width=\linewidth]{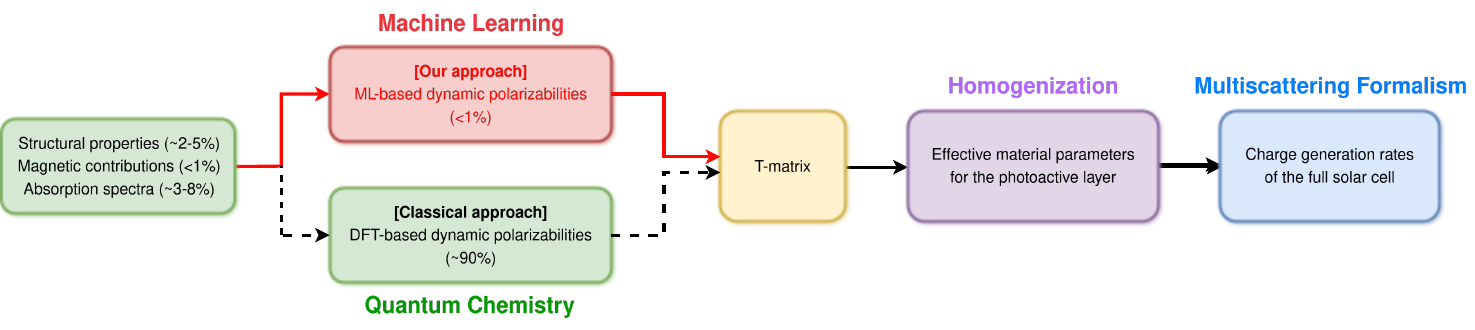}
    \caption{
        A scheme depicting the overall multiscale multiscattering simulations of the OPV stack. A neural network (NN) replaces computationally demanding calculations of dynamic polarizability tensors needed for the T-matrix construction step of the photoactive molecules.
    }
    \label{fig:scheme}
\end{figure}

The dynamic polarizability tensor forms the core input to the transition matrix (T-matrix) formalism~\cite{Waterman_Tmatrix, Fernandez-Corbaton:2020, Benedikt_multiscale}, which reformulates molecular optical responses into a representation directly compatible with Maxwell-based electromagnetic simulations. This pipeline propagates quantum-chemical information from the molecular scale to device-level optical models, enabling accurate prediction and optimization of OPV device performance (Figure~\ref{fig:scheme}). Recent work by some of the present authors has begun to realize fully differentiable digital twins of complete OPV device stacks, unifying material-, device-, and system-level simulations to enable end-to-end, gradient-based optimization of solar cell performance~\cite{Schubert2025-digitaltwin}. Extending this differentiability down to the molecular property calculation step itself, where conventional quantum-chemical methods such as time-dependent density functional theory (TD-DFT) are used, remains desirable and motivates machine-learning (ML) approaches capable of predicting these properties directly and differentiably.

Predictions of dynamic polarizability tensors typically rely on TD-DFT. Their computation remains a major bottleneck in molecular screening workflows. For example, calculating dynamic polarizability tensors for a subset of roughly 5\,000 molecules from the Quantum Machine 9 (QM9) dataset~\cite{QM9,QM9ramakrishnan2014quantum} requires approximately 13\,600 CPU hours. For the Harvard Organic Photovoltaic dataset (HOPV15)~\cite{Lopez2016-HarvardPhotoDataset}, which contains 347 larger and more structurally complex organic photoactive molecules relevant to OPV applications, the cost is around 60\,000 CPU hours. In contrast, computing only UV–vis spectra is far more efficient, requiring only about 500 CPU hours for the same QM9 subset and 5\,300 CPU hours for the HOPV15 molecules. UV–vis spectra are directly related to the imaginary part of the dynamic polarizability tensor, but can be obtained without resolving its full frequency-dependent, complex-valued structure. However, calculating dynamic polarizabilities is much more computationally complex than obtaining the discrete transitions that define the absorption spectrum. These numbers highlight that dynamic polarizability calculations dominate the computational budget in high-throughput OPV screening pipelines, accounting for roughly 80 to 90\% of the total cost.

Graph neural networks (GNNs) offer a promising alternative for predicting dynamic polarizability tensors at a fraction of the computational cost.
Message passing neural networks (MPNNs), introduced by \citeauthor{Gilmer2017}~\cite{Gilmer2017}, enabled end-to-end learning of molecular representations. Early architectures such as SchNet and DimeNet started incorporating geometric information to improve quantum-chemical predictions~\cite{Schutt2018_SchNet, Klicpera2020_DimeNet}. Equivariant architectures such as Tensor Field Networks~\cite{Thomas2018-tensorFieldNetworks}, SE(3)-Transformers~\cite{fuchs2020se3transformers3drototranslationequivariant}, EGNNs~\cite{satorras2022E_n}, and PaiNN~\cite{Kristof2021_PaiNN}, introduced rotation- and translation-equivariant message passing, enabling the learning of vector and tensor properties directly from atomic coordinates.

More recent and widely used models for molecular regression tasks include NequIP~\cite{Batzner2022-NequIP} and Allegro~\cite{Musaelian2023_Allegro}, which scale equivariant learning to large systems with high accuracy and data efficiency. Although these models internally propagate geometric information that can be viewed as low-order tensors, they primarily target scalar quantities such as energies (and, by extension, forces, which are the gradients of the energies).

Recent developments push beyond this, aiming to learn higher-order tensorial observables. DetaNet~\cite{Zou2023-DetaNet} introduces explicit higher-order tensor message passing to predict static polarizabilities and hyperpolarizabilities with quantum-chemical fidelity. HotPP~\cite{Wang2024-zn} follows a similar approach by passing Cartesian tensor embeddings to model vibrational and electronic response properties. Most recently, \citeauthor{Falletta2025-jf} proposed a unified equivariant framework in which response properties, including forces, polarization, and static polarizability, emerge as exact derivatives of a learned potential, guaranteeing physical symmetry preservation~\cite{Falletta2025-jf}.

Similarly, a recent $SO(3)$-equivariant local coordinate frame GNN proposed by Filling \textit{et al.}~\cite{filling2025direct} enables direct tensor regression but remains confined to predicting static molecular polarizability tensors from geometric structures on small-molecule datasets~\cite{Gonnheimer2026_MACEMDP, Wilkins2019_AlphaML, Yang2019_StaticPol, Nguyen2022_TensorialML}. Equivariant GNNs have also been applied to frequency-dependent dielectric tensors of crystals~\cite{Hsu2026_TensorialSpectra}. To the best of our knowledge, no existing equivariant model directly predicts frequency-dependent, complex-valued dynamic polarizability tensors from a molecular graph. 

We address this gap by extending the DetaNet architecture to incorporate both atomic coordinates and a sampled UV–vis spectrum. Although the model can be trained using atomic coordinates alone, adding spectral information substantially improves performance, and UV–vis spectra are comparatively inexpensive to obtain. This enables accurate, direct prediction of the full dynamic polarizability tensor across a range of frequencies, providing a fast, ML-based complement to TD-DFT. While contemporary frameworks focus on static properties of simpler structures, our model successfully scales to the larger, structurally complex molecules of the HOPV15 dataset. Furthermore, we validate its practical utility by passing these predicted polarizabilities into a realistic device-level optical modeling workflow, ultimately calculating macroscopic charge carrier generation rates for full organic photovoltaic solar cells.

We generate reference data using TD-DFT for two benchmark datasets: a sampled subset of the QM9 database, which we denote as QM9SPol, and the HOPV15 dataset for more application-relevant molecules. QM9SPol contains small organic molecules to benchmark and validate our methodology, while HOPV15 consists of larger and more structurally complex molecules closer to real-world applications.
We assess the model's performance on both QM9SPol and HOPV15 using various metrics, including the mean squared error (MSE), $R^{2}$ scores, and the earth mover’s distance (EMD).

To further evaluate the applicability of our model for realistic device-level simulations, we conduct a comprehensive optical analysis for a held-out subset of 35 application-relevant molecules from the HOPV15 database. Using both TD-DFT polarizabilities and our model predictions, we compute the corresponding molecular T-matrices and perform device-level optical simulations to obtain charge-carrier generation rates. The optical responses obtained from model-predicted polarizabilities show agreement ($R^2 = 0.94$) with those based on TD-DFT, demonstrating that our model serves as a promising alternative to TD-DFT for device-level optical simulations. This confirms that beyond accurately predicting dynamic polarizabilities, our approach also maintains the physical fidelity required for downstream modeling while reducing computational cost.

%% file: 2-methods.tex
\subsection{Datasets and data preparation}

In this study, we use two primary datasets. The QM9SPol dataset contains smaller organic molecules to benchmark and validate our methods, and the HOPV15 dataset contains larger organic molecules to assess performance in more application-relevant cases. The quantity we aim to predict is the polarizability tensor $\boldsymbol{\alpha}(\omega)$. The tensor describes how a molecule's electron cloud deforms in response to an external electric field oscillating at angular frequency $\omega$. It is a complex $3\times3$ tensor that describes the excitation of electric dipoles in a corresponding molecule. In the sections below, we explain how we calculate it for the different datasets and how we use the tensor's properties to train our model. 

\subsubsection*{The QM9SPol dataset}
The QM9SPol dataset (QM9 with UV–vis spectra and polarizabilities) is a sampled subset of the original QM9 dataset, containing 5\,160 randomly selected molecules from QM9~\cite{QM9, QM9ramakrishnan2014quantum}. We re-optimize all geometries, and we calculate static and dynamic polarizabilities, as well as the 15 lowest singlet excitation energies and the corresponding oscillator strengths. We compute dynamic polarizabilities across the 1.55–6.2 eV range (800 nm to 200 nm), at equally spaced 61 energies, including the full complex electric–electric (\textit{ee}), electric–magnetic (\textit{em}), and magnetic–magnetic (\textit{mm}) tensor components. Broadened absorption spectra based on oscillator strengths and dynamic polarizabilities can be related through the decadic absorption coefficient $\varepsilon(\tilde{\omega})$ in units of Lmol\textsuperscript{-1}cm\textsuperscript{-1}. The $\varepsilon(\tilde{\omega})$ calculated from oscillator strengths $f_j$ is: 
\begin{align}
\varepsilon(\tilde{\omega})=\frac{e^2N_A}{40\epsilon_{0}m_{e}c^{2}ln(10)}\sum_{j=1}^{n}{f_{j}L(\tilde{\omega};\tilde{\omega_{j}})}
\end{align}
where $e$ is the electron charge, $\epsilon_{0}$ is the vacuum permittivity, $N_A$ is the Avogadro constant, $m_e$ is the mass of the electron, $c$ is the speed of light, $\tilde{\omega}$ is the photon energy at which the spectrum is evaluated, $\tilde{\omega}_j$ is the excitation energy of the $j$-th singlet transition, and $L(\tilde{\omega};\tilde{\omega_{j}})$ is a Lorentzian function:
\begin{align}
L(\tilde{\omega};\tilde{\omega_{j}})=\frac{1}{\pi}\frac{\frac{1}{2}\tilde{\Gamma}}{(\tilde{\omega}-\tilde{\omega}_j)^2+( \frac{1}{2}\tilde{\Gamma})^2}
\end{align}
The $\tilde{\Gamma}$ is the full-width at half-maximum (FWHM).
The same values of $\varepsilon(\tilde{\omega})$ in units of Lmol\textsuperscript{-1}cm\textsuperscript{-1} can be calculated from dynamic polarizabilities $\alpha_{ee}$ provided in atomic units (a.u.) as: 
\begin{align}
\varepsilon(\tilde{\omega})=\frac{10a^{2}_{0}N_A}{ln(10)}\frac{4\pi\omega}{3c^{au}}Tr\{Im(\alpha_{ee}) \}
\end{align}
where $a_{0}$ is the Bohr radius and $c^{au}$ is the speed of light in vacuum in atomic units.
This relationship formalizes the physical link between UV–vis absorption spectra and the dynamic polarizability tensor, and motivates the use of UV–vis spectra as an auxiliary input to the model, as introduced above.
Finally, all dynamic polarizability tensors are interpolated to 241 energies covering the same energy range using the AAA algorithm~\cite{doi:10.1137/16M1106122, https://doi.org/10.1002/lpor.202400584, https://doi.org/10.1002/adts.202400989} to produce more detailed spectral features for training without substantial additional computational cost. In this work, we only consider the \textit{ee} component for the linear absorption. This dataset serves as an initial benchmark for our approach, as its molecules are small and structurally simple, allowing us to develop and validate the model before moving to more complex systems.

\subsubsection*{The HOPV15 dataset}
The HOPV15 dataset consists of 347 conjugated molecules and small polymers representative of acceptor materials in organic photovoltaic systems~\cite{Lopez2016-HarvardPhotoDataset}. Relative to the QM9SPol dataset, the molecular structures in HOPV15 exhibit greater chemical and structural complexity. For each molecule, we apply the same electronic structure protocol used for QM9SPol to compute the static polarizability at zero frequency and the frequency-dependent polarizability tensor components (\textit{ee}, \textit{em}, and \textit{mm}), along with the 40 lowest singlet excitation energies and their associated oscillator strengths. We use the same AAA interpolation procedure for both QM9SPol and HOPV15. HOPV15 is smaller than QM9SPol but more representative of realistic OPV molecules. Comprehensive details of all quantum chemistry calculations and their respective computational costs are available in the Supporting Information. 

\subsubsection*{Data preparation for ML training}
For each molecule, we subtract the static polarizability tensor, denoted by \( \boldsymbol{\alpha}(0) \in \mathbb{R}^{3\times 3} \), from the real part of every dynamic polarizability tensor. This centering step ensures that the model learns the relative spectral variations rather than absolute offsets, effectively removing molecule-specific baseline displacements.

The real part of the dynamic polarizability tensor is thus pre-processed as:
\begin{align}
\widetilde{\boldsymbol{\alpha}}'(\omega_f)
\;=\;
\boldsymbol{\alpha}'(\omega_f)
\;-\;
\boldsymbol{\alpha}(0),
\qquad
f=1,\dots,F.
\end{align}
Here, \( F \) denotes the number of frequencies considered in the dynamic spectra, \( \boldsymbol{\alpha}(\omega) \) denotes the frequency-dependent dynamic polarizability tensor, with \( \boldsymbol{\alpha}'(\omega) \) representing its real part and \( \boldsymbol{\alpha}''(\omega) \) its imaginary part. The static polarizability \( \boldsymbol{\alpha}(0) \) is equivalent to the dynamic polarizability evaluated at zero frequency.

After separating and centering, the real and imaginary parts are stacked as:
\begin{align}
\underbrace{\widetilde{\boldsymbol{\alpha}}'(\omega_1),\,\dots,\,\widetilde{\boldsymbol{\alpha}}'(\omega_{F})}_{\text{real part}}
\;\Bigl|\Bigr.\;
\underbrace{\boldsymbol{\alpha}''(\omega_1),\,\dots,\,\boldsymbol{\alpha}''(\omega_{F})}_{\text{imaginary part}},
\qquad
\boldsymbol{\alpha}^{(\cdot)}(\omega_f)\,\in\,\mathbb{R}^{3\times 3}.
\end{align}
After concatenation, we represent each sample as:
\begin{align}
\Bigl(\,\widetilde{\boldsymbol{\alpha}}'(\omega_{1:F}),\,\boldsymbol{\alpha}''(\omega_{1:F})\Bigr)
\;\in\;
\mathbb{R}^{2F\times3\times3}.
\end{align}
For each molecule in both datasets, we generate the simulated UV–vis spectra by broadening the calculated discrete oscillator strengths using Lorentzian line shapes with a FWHM of 0.05\,eV. The oscillator strength quantifies the transition probability from the electronic ground state to an excited singlet state, and together with the corresponding excitation energy, determines the contribution of each dipole transition to the absorption spectrum.

We compute the UV–vis intensity at a given photon energy $E$ using the following expression:
\begin{align}
I_{\text{Lorentz}}(E)=\sum_n \frac{f_{0\rightarrow n}}{1+\bigl[(E - E_{0\rightarrow n})/\gamma\bigr]^2},
\end{align}
where $f_{0\rightarrow n}$ is the oscillator strength of the transition from the ground state to the $n$-th excited state, $E_{0\rightarrow n}$ is the corresponding vertical excitation energy, and $\gamma = \text{FWHM}/2 = 0.025\,\text{eV}$ is the Lorentzian half-width at half-maximum. $I_{\text{Lorentz}}(E)$ is the resulting absorption intensity at energy $E$, and the summation runs over all calculated singlet excited states.

We apply the Lorentzian broadening directly to the oscillator strengths, and sample the resulting continuous spectra at the same photon energies used for the dynamic polarizability tensors. This ensures consistent input representation across both spectral modalities.

\begin{figure}[!htb]
    \centering
    \includegraphics[width=\linewidth]{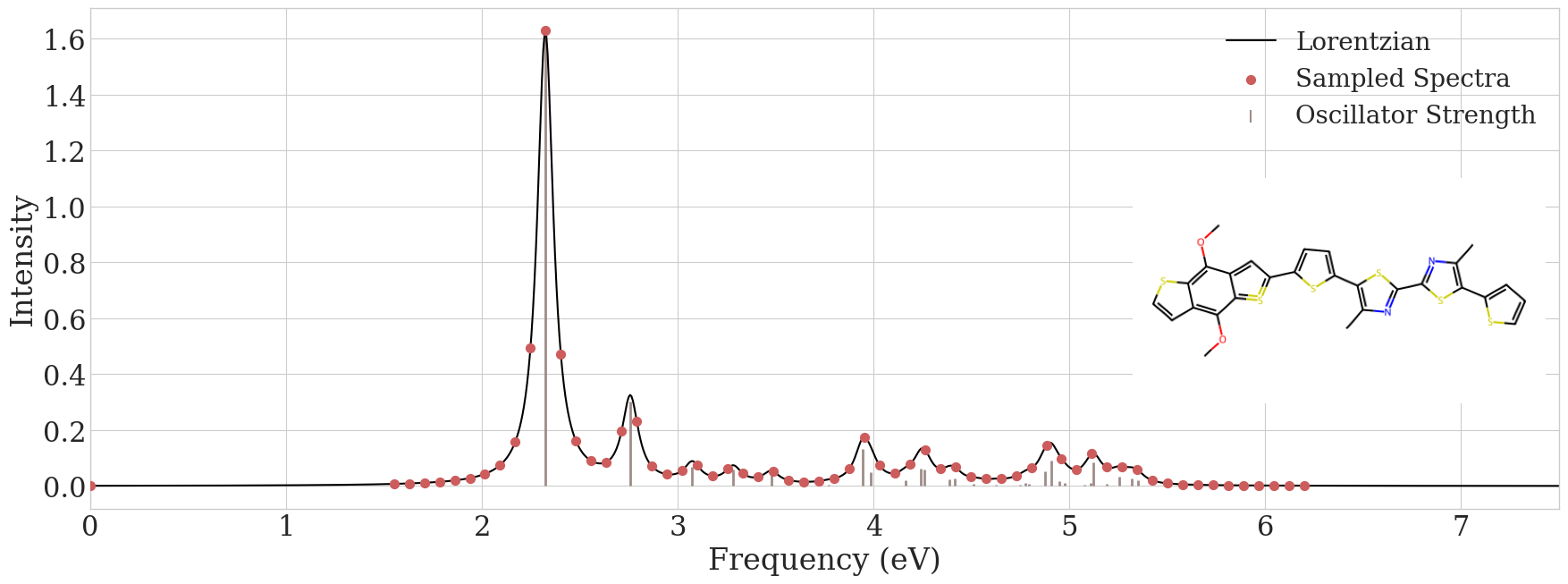}
    \caption{
        Sampled UV–vis spectra generation and visualization. 
        The solid black line shows the computed Lorentzian line shapes corresponding to the lowest singlet excitation energies and oscillator strengths, with a FWHM of 0.05\,eV. 
        Red dots represent the UV–vis intensities sampled at the same frequencies as the dynamic polarizability tensors. 
        The vertical gray bars indicate the positions and relative intensities of the raw oscillator strengths prior to broadening. 
        To the right, the corresponding molecular structure is displayed.
    }
    \label{fig:sampled_uv_vis_spectra}
\end{figure}

We illustrate the construction of the UV–vis spectra in Figure~\ref{fig:sampled_uv_vis_spectra}.

\subsection{Model architecture}
\begin{figure}[!htb]
    \centering
    \includegraphics[width=\linewidth]{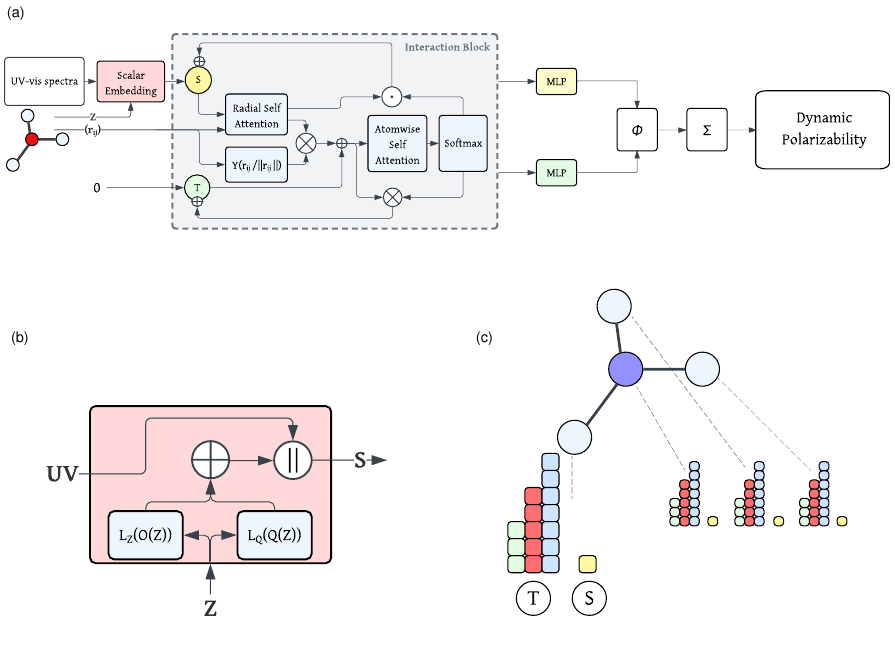}
    \caption{
        Overview of the model architecture and key submodules.
        (a) Overview of the model architecture, from the molecular geometry and UV–vis spectrum on the input side, through the interaction block (dashed box), to the global embedding and the predicted dynamic polarizability tensor on the output side.
        (b) The scalar embedding block, combining atomic-number information with the sampled UV–vis spectrum into the initial scalar embedding \(S^0\).
        (c) Irreducible representation structure used in the model, with colors indicating features of a given rotation degree \(\ell\) and parity \(p\).
    }
    \label{fig:model_overview}
    \label{fig:architecture}
    \label{fig:embedding}
    \label{fig:irreps}
\end{figure}

Our approach is grounded in the DetaNet framework but incorporates extensions to process UV–vis spectra alongside atomistic information. The model takes as input the 3D molecular geometry, atomic numbers, and the UV–vis spectrum evaluated at the same frequencies as the dynamic polarizability tensors. We represent the overall architecture in Figure~\ref{fig:architecture}.

Scalar features \(S_i^n\) and irreducible tensor features \(T_{i,\ell,p}^{\,n,m}\) are processed separately. We visualize the features per atom in Figure \ref{fig:irreps}. 

While we initialize tensor features to zero, we construct the initial scalar block \(S_i^0\) from a nuclear one-hot encoding \(O(Z_i)\) and a handcrafted electronic descriptor \(Q(Z_i)\), as shown in Figure~\ref{fig:embedding} and as given by:
\begin{align}\label{eq: S_i^0}
(S^{0}_{i}=L_{\text{emb}}\!\bigl(L_{Z}(O(Z_i))+L_{Q}(Q(Z_i))\bigr)\;\Vert\;UV)
\end{align}
The symbol \(\Vert\) denotes feature-wise concatenation, where \(UV\) represents the sampled UV–vis spectrum and $L$ denotes learnable linear layers.

Within the interaction block, the model first applies radial self-attention over neighboring atom pairs, weighted by their interatomic distance, to construct distance-dependent edge messages and encode geometric information via real spherical harmonics \(Y_{\ell m}(\hat{\mathbf{r}}_{ij})\). Tensor products (\(\times\)), dot products (\(\cdot\)), and standard additions (\(+\)) are used throughout to combine features while maintaining equivariance. Atomwise self-attention modules then update each atom's own aggregated scalar and tensorial features, refining the representation built from the preceding neighbor messages. The model maintains rotational and permutational equivariance throughout this process.

After \( N \) iterations, the updated scalar and tensorial features are passed through two multi-layer perceptrons (MLP), ultimately producing the dynamic polarizability tensor at each frequency point in a single forward pass. At the output stage, the model assembles the frequency-resolved dynamic polarizability tensor via a summation operator (\(\Sigma\)) and a property prediction head (\(\Phi\)), which combines learned features.

To handle second-order tensorial properties such as dynamic polarizabilities \( \boldsymbol{\alpha}^{(f)} \), the model uses a group-theoretical decomposition into scalar and traceless parts:
\begin{align}
\bigl\{\boldsymbol{\alpha}^{(f)}\bigr\}_{f=1}^{F} 
= 
\bigl\{s_b^{(f)} \;\oplus\; \bigl(t_{2,1}^{(f)} + s_a^{(f)} Y^{(2e)}(\hat{\mathbf r}_{i})\bigr)\bigr\}_{f=1}^{F}.
\end{align}

Here, \(s_b^{(f)}\) represents the scalar (trace) component, while \(s_a^{(f)}\) is a learnable scaling factor that modulates the non-local traceless contribution encoded by the spherical harmonics \(Y^{(2e)}(\hat{\mathbf r}_{i})\), of rotation degree \(\ell=2\) and even parity (\(p=1\)). The term \(t_{2,1}^{(f)}\), likewise of degree \(\ell=2\) and even parity, captures local contributions to the traceless tensor. By predicting the entire set \(\{\boldsymbol{\alpha}^{(f)}\}\) simultaneously, the model captures the frequency-dependent structure of the dynamic polarizability tensor in a single forward pass, ensuring consistent and efficient learning of both local and global directional dependencies.

Because all submodules, including tensor products and reduction steps from \textit{e3nn}~\cite{geiger2022e3nneuclideanneuralnetworks} (the Euclidean neural network library implementing equivariant operations on irreducible representations) are implemented in an equivariant manner, the model fully preserves physical symmetry constraints throughout. Nonlinearities and biases are restricted to the scalar channels, ensuring that higher-order tensor features remain equivariant by design.

\subsection{Post-polarizability prediction optical simulation pipeline}

To compare model-predicted dynamic polarizabilities with their TD-DFT counterparts in a realistic device context, we integrate both into the same optical simulation workflow. For this purpose, we first convert the dynamic polarizabilities into molecular T-matrices~\cite{Waterman_Tmatrix, Fernandez-Corbaton:2020, T_matrix_polarizabilities_Ivan}, which describe light–matter interactions at the single-molecule level~\cite{Benedikt_multiscale, Benedikt_homogenization}. We manually set predicted negative, diagonal imaginary dynamic polarizability values to zero prior to this step. These negative values occur only occasionally as prediction errors of the ML model; clipping them to zero corrects unphysical artifacts without affecting any TD-DFT-based results. We homogenize the resulting T-matrices using the \textit{treams} framework to obtain the effective bulk T-matrix and corresponding permittivity of the photoactive layer~\cite{treams_Beutel}.

Using these material properties, we carry out device-level optical simulations with a semi-analytical scattering-matrix (S-matrix) solver, assuming the solar cell is a stack of stratified isotropic media~\cite{Sax_Github}. The S-matrix is a linear operator connecting the incoming and outgoing waves of the stack~\cite{S_matrix_Cotter}. It is constructed analytically for each layer from its permittivity $\varepsilon$ and thickness $d$~\cite{S_matrix_STEFANOU199849}, and individual layers are successively combined via the Redheffer product~\cite{Redhaffer_star_product, Redheffer_product_original} to obtain the overall transmitted field $\mathrm{E}(x)$. From this field, we calculate the generation rate $G$ of the stack as:
\begin{equation} \label{eq:gen_rate}
G = \mathrm{e} \int_{x}
    \int_{\lambda} P_{\mathrm{abs}}(x)\,
    f(\lambda)\, \mathrm{d}\lambda\, \mathrm{d}x,
\end{equation}
with $\mathrm{e}$ the elementary charge and factor $f(\lambda)$ defined as $f(\lambda) = \mathrm{Irr}(\lambda)\, \frac{P_\mathrm{in}}{A_\mathrm{in}}$, where $\mathrm{Irr}$ is the irradiance at wavelength $\lambda$ based on the standard AM 1.5 spectrum, $P_\mathrm{in}$ is the input power over the illuminated area $A_\mathrm{in}$, and $P_{\mathrm{abs}}(x)$ is the spatially resolved absorption of the considered layers, which depends on the resulting field $\mathrm{E}(x)$.

Running the entire pipeline described above for both TD-DFT and model-based polarizabilities allows a direct, device-level comparison of the two approaches within a realistic OPV workflow.

%% file: 3-results_discussion.tex
\subsection{Evaluation metrics of the model}

Commonly used metrics for the prediction of molecular properties, such as the static polarizability, are the $R^2$ score, MSE or root mean squared error (RMSE), and mean absolute error (MAE)~\cite{Kristof2021_PaiNN, Zou2023-DetaNet, Wang2024-zn, Grisafi2018-df}. These are all pointwise metrics, calculated via a direct element-by-element comparison between predicted and reference values, averaged over all tensor components and frequency points.

However, since our goal is to predict the full frequency-dependent polarizability tensor, pointwise metrics are insufficient. Therefore, we adopt the EMD, or 1-Wasserstein distance, which quantifies discrepancies in the overall spectral shape of each tensor component across frequencies, providing a more meaningful evaluation for spectrum-level predictions.
The EMD conceptualizes each spectrum as an area distributed across frequency. The distance measures the minimal total cost required to redistribute the predicted spectrum to match the target spectrum, with cost defined as the product of the total area moved and the distance it is moved. Unlike pointwise metrics, the EMD captures differences in the overall shape of the spectrum and is less sensitive to small shifts of sharp peaks, which can otherwise dominate error measures like the RMSE.

For each matrix element \( i \in \{1,\dots,18\} \) from the real and imaginary parts of dynamic polarizability, the distributions of predicted and reference values, denoted as \(\mu^{(i)}\) and \(\nu^{(i)}\), are compared using the 1-Wasserstein distance \( W_1(\mu^{(i)}, \nu^{(i)}) \). The overall EMD loss for every polarizability tensor across frequencies is then defined as:
\begin{align}
\mathcal{L}_{\mathrm{EMD}} = \frac{1}{18} \sum_{i=1}^{18} \min_{\gamma \in \Gamma(a,b)} \langle \gamma, \mathbf{M}^{(i)} \rangle_F,
\end{align}
where \(\mathbf{M}^{(i)}\) is the ground-cost matrix of pairwise Euclidean distances computed between the predicted and target values across all frequencies for component \(i\). The notation \(\langle \cdot, \cdot \rangle_F\) denotes the Frobenius inner product, which sums the element-wise products of the matrices. The transport plan \(\gamma\) satisfies the marginal constraints \(a = b = \frac{1}{B}\), ensuring mass conservation. In practice, the optimal transport cost is computed using the network simplex algorithm implemented in the Python Optimal Transport (POT) library \cite{POT}.

\subsection{Training results}

Prior to training, we randomly partition both datasets into training and test sets using a 90/10 split. For QM9SPol, this results in 4644 molecules for the training set and 516 for the test set. For HOPV15, the split yields 312 molecules for training and 35 for the test set. We decompose the dynamic polarizability tensor, sampled at \( F=61 \) frequency points, into its real and imaginary components and concatenate it to form a tensor of size \( 122 \times 3 \times 3 \). For HOPV15, which consists of photoactive molecules relevant to the target application, we train three separate models: one predicting 61 polarizability values, one fine-tuned from a model pre-trained on QM9SPol (also 61 values), and one predicting 241 polarizability values, obtained by expanding the original 61 sampled frequencies with additional points generated via the AAA algorithm.
For the UV–vis spectra input, we choose to sample 61 points to match the frequencies used in the polarizability tensors. Although higher sampling densities are tested, only a slight performance gain is observed. We also evaluate models trained solely on atomic coordinates. However, these baselines consistently show higher error across both QM9SPol and HOPV15, motivating the inclusion of UV–vis spectra, which are comparatively inexpensive to obtain.

We optimize model performance through Bayesian hyperparameter tuning, and apply a dropout rate of 20\% to all hidden layers to prevent overfitting. We summarize the performance of the models across different experiments in Table~\ref{tab:deta_results}. Although the EMD is the more relevant metric, we also include the RMSE and $R^2$ scores to make further comparisons. 

\begin{figure}[htbp]
    \centering
    \includegraphics[width=\linewidth]{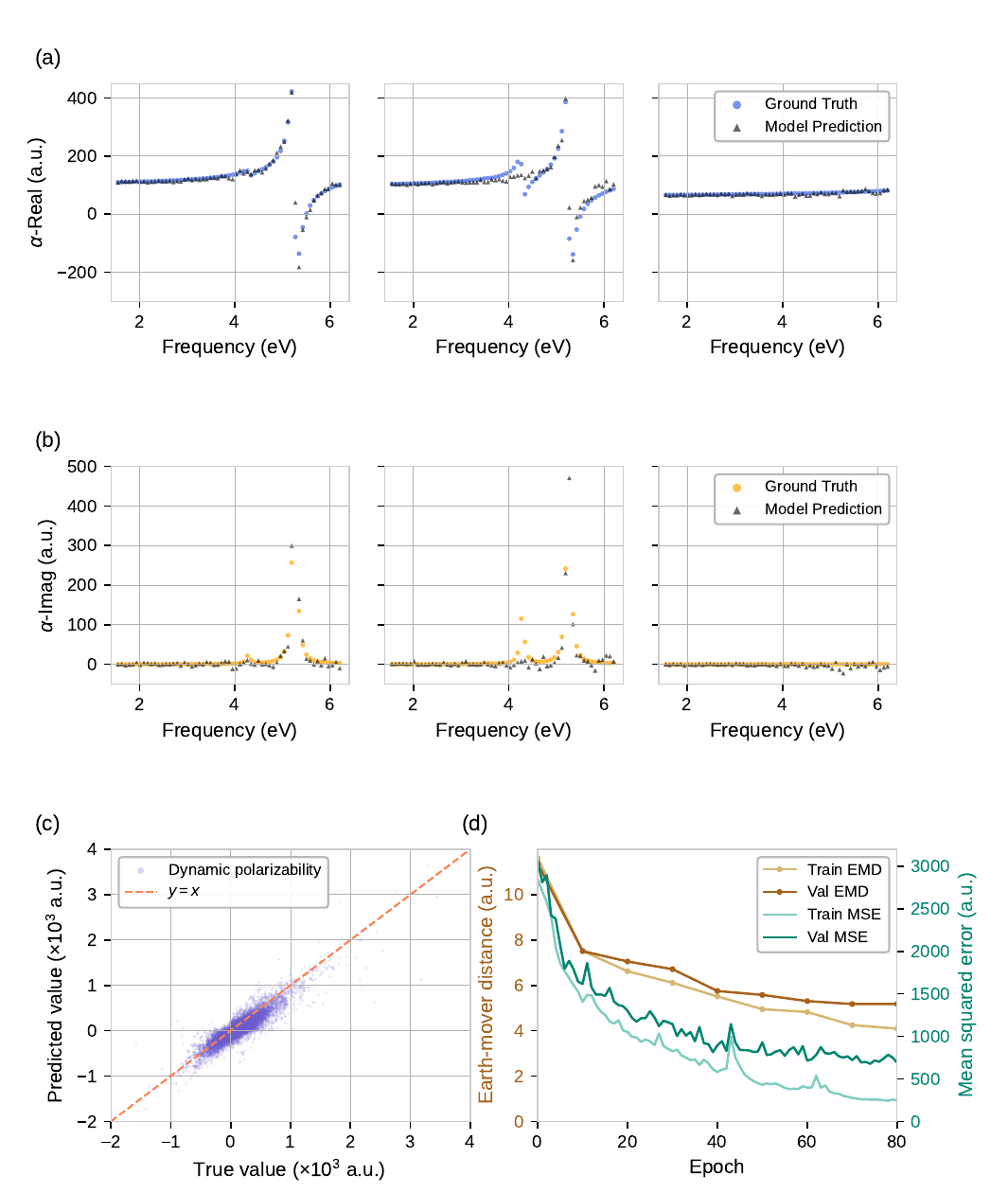}
    \caption{\textbf{QM9SPol:} (a) Real and (b) imaginary parts of the trace components of the dynamic polarizability tensor elements (61 polarizabilities) of a single molecule from the QM9SPol dataset, shown for the \textit{xx}, \textit{yy} and \textit{zz} components (left to right in each row). Blue and yellow circles denote ground-truth (GT) values; black triangles show model predictions. (c) Point-wise comparison of predicted versus ground-truth values across the test set. (d) Training history: EMD (left axis) and MSE (right axis) for both training and validation splits over 80 epochs.}
    \label{fig:polar_overview_qm9spol}
\end{figure}

For the QM9SPol dataset, the model effectively captures the overall shape of the frequency-dependent dynamic polarizability tensor, although it occasionally misses minor peak structures. These discrepancies may originate from the relatively limited size of the dataset, which could lead to overfitting, especially considering that DetaNet was originally trained on over 130\,000 molecules to achieve accurate UV–vis spectrum predictions. Predicting the dynamic polarizability tensor across the 1.55--6.2~eV frequency range is a more demanding task and likely requires a substantially larger and more diverse dataset for optimal performance. We illustrate the model’s performance on QM9SPol in Figure~\ref{fig:polar_overview_qm9spol}.

Fine-tuning from QM9SPol to HOPV15 does not offer an advantage over training the model on HOPV15 from scratch. Molecules in the HOPV15 dataset are larger, structurally more complex, and differ substantially in their physical properties from those in QM9SPol, limiting the effectiveness of knowledge transfer between the two datasets.

\begin{figure}[htbp]
    \centering
    \includegraphics[width=\linewidth]{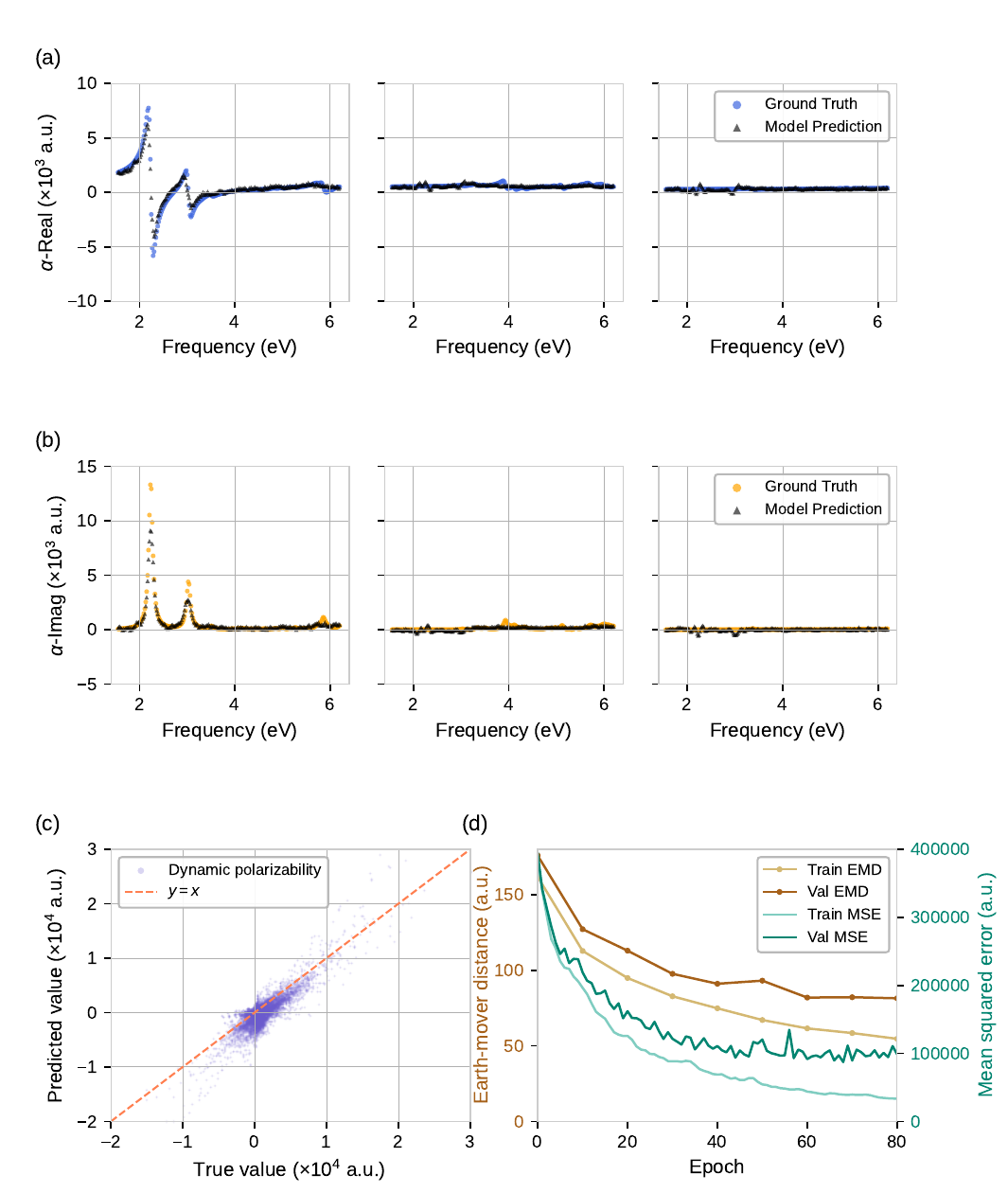}
    \caption{\textbf{HOPV15:} (a) Real and (b) imaginary parts of the trace components of the dynamic polarizability tensor elements (241 polarizabilities) of a single molecule from the HOPV15 dataset, shown for the \textit{xx}, \textit{yy} and \textit{zz} components (left to right in each row). Blue and yellow circles denote ground-truth values; black triangles show model predictions. (c) Point-wise comparison of predicted versus ground-truth values across the test set. (d) Training history: EMD (left axis) and MSE (right axis) for both training and validation splits over 80 epochs.}
    \label{fig:polar_overview_HOPV241}
\end{figure}

On the HOPV15 dataset, the model effectively learns the structure and variation of the dynamic polarizability tensor across the frequency range.
The models trained on 61 and 241 polarizabilities produce comparable results, with the 241-point model performing slightly better on RMSE and EMD, though with a marginally lower $R^2$ score. The finer sampling grid improves the model’s ability to capture smaller peaks in the polarizability tensor. We show an overview of the best-performing 241-point model in Figure~\ref{fig:polar_overview_HOPV241}.

\begin{table}[htbp]
    \centering
    \caption{Performance comparison of models trained on QM9SPol and HOPV15. The table indicates whether UV–vis spectra are included as input and the number of sampled frequency points used for each model. It also includes results from a fine-tuning experiment where a QM9SPol-pretrained model was adapted to HOPV15.}
    \label{tab:deta_results}
    \small
    \setlength{\tabcolsep}{4pt}
    \begin{tabular}{@{}lccccc@{}}
        \toprule
        Dataset (Training Regime) & UV–vis Input & Frequency points & RMSE (a.u.) & EMD (a.u.) & $R^2$ \\
        \midrule
        QM9SPol & \ding{55} & 61 & 48.7 & 8.5 & 0.55 \\ 
        QM9SPol & \ding{51} & 61 & \textbf{22.1} & \textbf{5.8} & \textbf{0.86} \\ 
        \midrule
        HOPV15 & \ding{55} & 61 & 586.6 & 120.0 & 0.30 \\
        HOPV15 & \ding{51} & 61 & 261.9 & 80.3 & \textbf{0.82} \\
        HOPV15 & \ding{51} & 241 & \textbf{231.6} & \textbf{71.7} & 0.79 \\
        \midrule
        QM9SPol $\rightarrow$ HOPV15 (fine-tuning) & \ding{51} & 61 & 332.6 & 91.7 & 0.75 \\
        \bottomrule
    \end{tabular}
\end{table}

\subsection{Integration of the model into optical device simulations}

To understand how our model performs in a realistic device scenario, we apply it to an OPV solar cell stack. We consider 35 molecules of the HOPV dataset, which make up the test set. We introduce these molecules as acceptor molecules in the photoactive material of the solar cell. We then compare the charge-generation rates calculated with a computational pipeline described below, using the original TD-DFT-calculated dynamic polarizability tensors versus those using the predicted dynamic polarizability tensors. 

The OPV solar cell stack considered for this use case is depicted in Figure~5 of the SI. The sunlight enters the stack through a glass substrate with a thickness of \qty{1}{\micro\meter}. The next layer is a transparent indium tin oxide (ITO) electrode of \qty{150}{\nano\meter} thickness. We chose zinc oxide (ZnO) with a thickness of \qty{30}{\nano\meter} as the electron transport layer. The next layer is the photoactive layer with a thickness of \qty{200}{\nano\meter}. The hole transport layer is made of poly(3,4-ethylenedioxythiophene):poly(styrene sulfonate) (PEDOT:PSS) with \qty{190}{\nano\meter} thickness. The PEDOT:PSS layer is interrupted by a semi-transparent back contact of silver nanowires, which is approximated by a silver (Ag) slab of \qty{10}{\nano\meter} thickness. 

We define the photoactive material as a 1:1 volume blend of PM6 and a HOPV molecule from the test set. Poly[(2,6-(4,8-bis(5-(2-ethylhexyl-3-fluoro)thiophen-2-yl)-benzo[1,2-b:4,5-b’]dithiophene))-alt-(5,5-(1’,3’-di-2-thienyl-5’,7’-bis(2-ethylhexyl)benzo[1’,2’-c:4’,5’-c’]dithiophene-4,8-dione)] (in short, \textbf{PM6}) is a polymer, which primarily absorbs in the visible spectrum and acts as a donor polymer. The HOPV molecule, meanwhile, acts as the acceptor molecule. Further details are given in the Supporting Information.

For each PM6:HOPV combination, we calculate the generation rate of the stack (Eq.~\ref{eq:gen_rate}) using the scattering-matrix pipeline described in Methods, which requires the permittivity values of the chosen layers, especially of the photoactive material blend.

For each HOPV molecule, we convert dynamic polarizabilities obtained from model predictions and from TD-DFT calculations into molecular T-matrices and homogenize these (as described in the Methods section), yielding the effective permittivity \textrm{$\bm{\varepsilon}_{\textrm{eff}}$} and permeability \textrm{$\bm{\mu}_{\textrm{eff}}$} of the photoactive material~\cite{Benedikt_homogenization}. The permittivity values of the bulk HOPV material are then combined in a 1:1 volume ratio with the PM6 permittivity values, known from separate TD-DFT calculations. For the other layers of the solar cell stack, known permittivity values were used. Further theoretical details are given in the Supporting Information.

\begin{figure}
    \centering
    \includegraphics[width=0.65\textwidth]{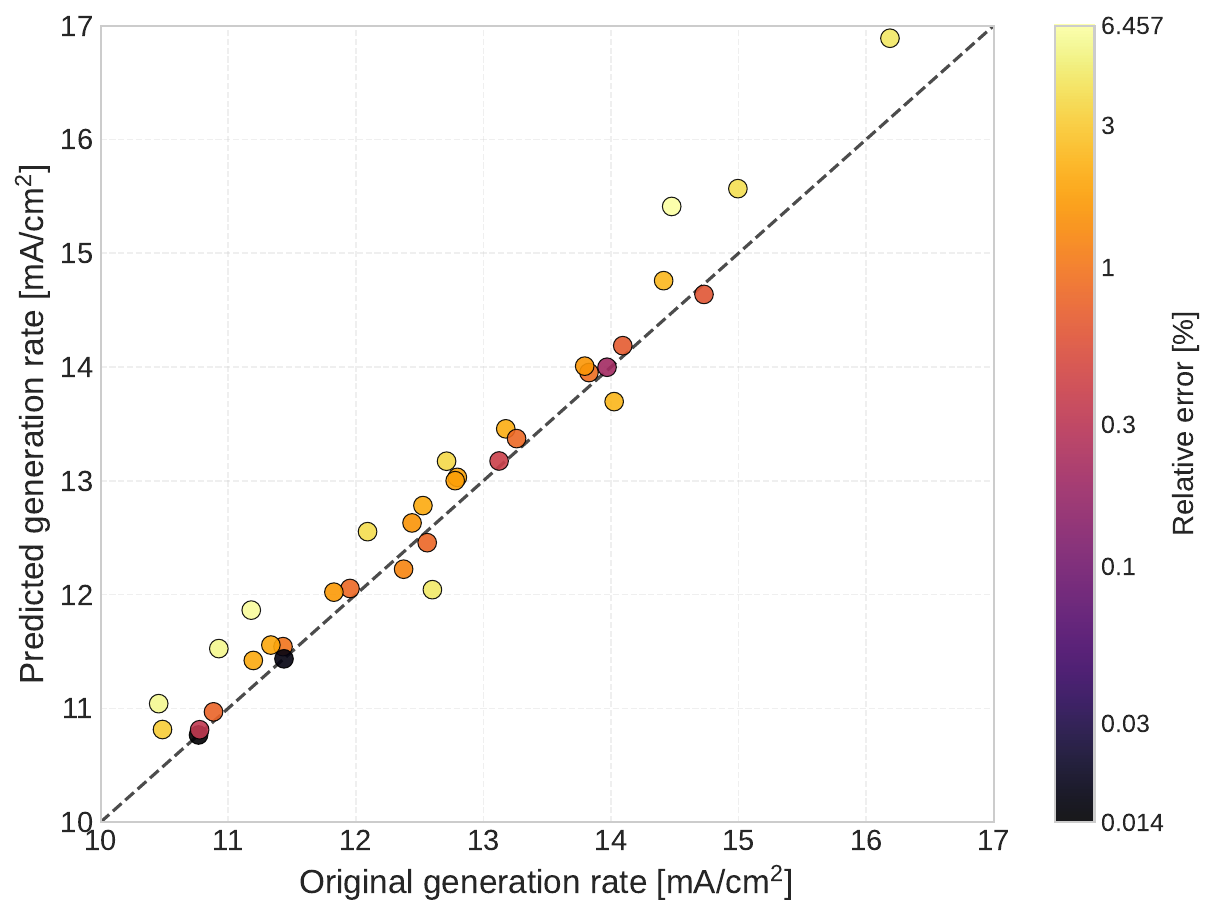}
    \caption{Calculated generation rate \textrm{GR$_{\textrm{pred}}$} plotted over \textrm{GR$_{\textrm{original}}$} in units of $[\mathrm{mA}/\mathrm{cm}^2]$. The predicted generation rate \textrm{GR$_{\textrm{pred}}$} is based on dynamic polarizabilities predicted by our machine-learning model. The original generation rate \textrm{GR$_{\textrm{original}}$} is based on the original dynamic polarizabilities calculated with TD-DFT. The colors of the dots indicate the relative error rate. The relative errors range from 0.014\% to 6.457\%. The dashed line marks $y=x$. Metrics over the 35 HOPV molecules: $R^2 = 0.9367$, RMSE = 0.3590 $\mathrm{mA}/\mathrm{cm}^2$.}
    \label{fig:generation_rates}
\end{figure}

Finally, we obtain one predicted generation rate \textrm{GR$_{\textrm{pred}}$} (based on the predicted dynamic polarizabilities of our model) and one ground-truth, original generation rate \textrm{GR$_{\textrm{original}}$} (based on the dynamic polarizabilities directly calculated with TD-DFT) for every considered PM6:HOPV combination. To compare the two types of generation rate, we calculate general metrics such as the $R^2$ score and RMSE, as well as the relative error between \textrm{GR$_{\textrm{pred}}$} and \textrm{GR$_{\textrm{original}}$} (Figure~\ref{fig:generation_rates}). The overall metrics that were reached are an $R^2$ score of 0.9367 and an RMSE of 0.3590 $\mathrm{mA}/\mathrm{cm}^2$, and the relative errors range from 0.014\% to 6.457\%, indicating that generation rates derived from predicted dynamic polarizabilities based on our model are reliable for downstream generation rate calculations in an OPV application.

%% file: 4-conclusion_outlook.tex
In this work, we present an equivariant machine-learning framework to predict frequency-dependent dynamic polarizability tensors for complex organic molecules.
By integrating molecule geometries with inexpensive UV–vis spectral data, our model captures the intricate spectral variations essential for optical modeling. The primary strength of this approach lies in its integration into a multi-scale simulation pipeline: by feeding model-predicted polarizabilities into an OPV device-level optical simulation, we obtained charge carrier generation rates in agreement with TD-DFT-based results ($R^2 = 0.9367$), directly demonstrating the practical utility of our approach beyond isolated polarizability prediction.
Ultimately, our approach provides a computationally efficient complement, or even an alternative, to expensive TD-DFT workflows. By enabling the rapid screening of photoactive materials directly within a device-level context, this framework paves the way for the accelerated discovery and optimization of next-generation organic solar cells.
Moreover, because our model is differentiable with respect to the input molecular geometry, it supports gradient-based optimization workflows in which molecular geometries can be directly refined. This offers a scalable path toward the automated discovery of high-performance materials for organic photovoltaics.
Looking further ahead, this differentiable polarizability model is well suited to serve as one module within a fully differentiable digital twin of the OPV stack, unifying material-, device-, and system-level simulations into a single end-to-end optimizable pipeline, as outlined in previous work~\cite{Schubert2025-digitaltwin}. Embedding the present model into such a framework would allow gradients to propagate all the way from molecular structure to predicted energy yield, enabling fully automated, physics-informed optimization of photoactive materials for next-generation organic photovoltaics.

%% file: si.tex
\clearpage
\setcounter{figure}{0}
\setcounter{table}{0}
\renewcommand{\thefigure}{\arabic{figure}}
\renewcommand{\thetable}{\arabic{table}}

\section*{Supporting information}

\subsection*{Theoretical background}

\subsubsection*{Time-dependent density functional theory}
\label{sec:td-dft}
Density Functional Theory (DFT) is a practical way to describe how atoms and molecules behave.
An atom is a mostly empty space: almost all of its mass sits in the nucleus, yet its electrons decide how
the atom reacts with its surroundings. In principle, one could predict the motions of the electrons by solving Schrödinger's equation, but that task becomes impossibly hard as soon as many atoms are involved because every electron interacts with every other one~\cite{video_TDDFT}.

DFT sidesteps this problem by focusing on the electron density, the probability of finding an electron at any point in space, instead of tracking each electron separately.
Working with this density turns the many-body problem into a set of much simpler equations that computers can solve approximately, even for molecules containing hundreds or thousands of atoms, well beyond the reach of more accurate but far computationally expensive wavefunction-based methods such as coupled-cluster (CC) or configuration-interaction (CI)~\cite{video_TDDFT}.

When a molecule is placed in a time-varying electric or magnetic field, its electron cloud adjusts accordingly. \emph{Time-Dependent} Density Functional Theory provides a systematic way to predict that adjustment: it treats the external field as a small perturbation of the ground state calculated by DFT and, from the resulting linear response, extracts molecular properties such as the dynamic polarizability.

\subsubsection*{Static and dynamic polarizability}
\label{sec:polarizability}
When a molecule is placed in an electric field, the positively charged nuclei and the negatively charged electrons are displaced in opposite directions, creating an induced dipole moment $\mathbf p_{e}$. The polarizability $\boldsymbol{\alpha}$ is the proportionality factor between this dipole and the applied field~\cite{2023_LibreTextChemie}:
\begin{equation}
\mathbf p_{e}= \boldsymbol{\alpha}\,\mathbf E .
\end{equation}

For an isotropic, perfectly spherical molecule $\boldsymbol{\alpha}$ is just a scalar, but real molecules are usually anisotropic, and the response depends on the field direction. In that case, $\boldsymbol{\alpha}$
becomes a $3\times 3$ tensor.
\[
\boldsymbol{\alpha} =
\begin{pmatrix}
\alpha_{xx} & \alpha_{xy} & \alpha_{xz}\\
\alpha_{yx} & \alpha_{yy} & \alpha_{yz}\\
\alpha_{zx} & \alpha_{zy} & \alpha_{zz}
\end{pmatrix},
\]
so that each Cartesian component of $\mathbf E$ can induce a dipole along any axis.

\paragraph{Static vs.\ dynamic}
If the external field is constant, the corresponding \textit{static} polarizability is denoted $\boldsymbol{\alpha}(0)$.
When the field oscillates with an angular frequency $\omega$, the electron cloud cannot follow instantaneously, and the response becomes frequency-dependent, giving the \textit{dynamic} polarizability
$\boldsymbol{\alpha}(\omega)$. In general,
\[
\boldsymbol{\alpha}(\omega)=
\boldsymbol{\alpha}'(\omega)+
i\,\boldsymbol{\alpha}''(\omega),
\]
where the real part $\boldsymbol{\alpha}'$ describes the distortion of the charge cloud, while the imaginary part $\boldsymbol{\alpha}''$ quantifies the component associated with the absorption of electromagnetic energy when $\omega$ is close to an electronic resonance.

\subsubsection*{Computation of the polarizability tensors}
\label{sec:polar}

When a molecule is illuminated by an electromagnetic field $\{\mathbf E(\omega),\mathbf H(\omega)\}$, the induced electric and magnetic dipole moments $p_{e}$, $p_{m}$ vary linearly with the applied electric and magnetic fields:
\begin{align}
\mathbf p_{e}(\omega) &=\;
      \underbrace{\alpha_{ee}(\omega)}_{\text{electric--electric}}\,
      \mathbf E(\omega)\;
   +  \underbrace{\alpha_{em}(\omega)}_{\text{electric--magnetic}}
      \mathbf H(\omega),
      \label{eq:si_pe}\\[4pt]
\mathbf p_{m}(\omega) &=\;
      \underbrace{\alpha_{me}(\omega)}_{\text{magnetic--electric}}
      \mathbf E(\omega)\;
   +  \underbrace{\alpha_{mm}(\omega)}_{\text{magnetic--magnetic}}
      \mathbf H(\omega).
      \label{eq:si_pm}
\end{align}
Each $\alpha_{\nu\nu'}(\omega)$ is a $3\times3$ complex polarizability tensor that links a field component $\nu'\!\in\{e,m\}$ to the induced dipole component $\nu\!\in\{e,m\}$~\cite{Zerulla2024_Diss}.

To obtain these tensors, one adds a small, time-dependent perturbation to the ground-state Kohn--Sham Hamiltonian and solves the resulting coupled-perturbed Kohn--Sham equations.
While the frequency-dependent polarizability tensor $\alpha_{\nu\nu'}(\omega)$ is often introduced as a sum over transition dipole moments between the ground and excited states, it is more commonly calculated as the second derivative of the total molecular energy with respect to external perturbation~\cite{Zerulla2024_Diss}.
Repeating the procedure independently for each frequency provides the full \emph{dynamic} polarizability tensor required for subsequent optical simulations~\cite{Zerulla2024_Diss}.

\subsection*{Integration of the model into organic photovoltaic device simulations}

In this section, we provide a detailed description of the computational setup for all steps, starting with the molecular description of the single molecules, over the T-matrix construction and the prediction with the machine learning model, and finally to the calculation of the generation rates of the whole stacked organic photovoltaic device using both calculated and predicted dynamic polarizabilities. This quantum-chemistry-to-device pipeline forms the basis for the fully differentiable digital twin of the OPV stack described in previous work~\cite{Schubert2025-digitaltwin}. All produced quantum chemistry outputs from TURBOMOLE are deposited in the NOMAD database and can be accessed via the following DOI: \url{https://doi.org/10.17172/NOMAD.RG5M-AQT6}.

\subsubsection*{Details of the quantum chemistry calculations of QM9SPol and HOPV15 sets}

All quantum chemistry calculations in this work have been performed with the development version of the electronic structure program TURBOMOLE~\cite{TURBOMOLE2023, TM_today_tomorrow} that is developed at our university, the Karlsruhe Institute of Technology. For the ground state properties we used DFT, and for the excited state properties we employed its time-dependent linear response variant, TD-DFT.

\paragraph*{QM9SPol set} \label{subsec:si_QM9SPol}

A subset of 5160 molecules was randomly selected from the QM9~\cite{QM9,QM9ramakrishnan2014quantum} dataset containing over 134000 molecules with up to 9 heavy (not hydrogen) atoms. 

Atomic positions of all molecules were optimized using the gradient minimization technique to find stable ground state configurations using computationally cheap generalized-gradient approximation (GGA) PBE exchange-correlation (XC) energy functional \cite{PBE_a, PBE_b} with Karlsruhe atomic orbital (AO) split-valence basis set def2-SVP \cite{weigend2005a}. The minimum of the molecular structure on the potential energy surface was confirmed by performing vibrational frequency analysis. In order to further speed up all calculations, a resolution-of-identity (RI)\cite{ri}, and multipole-accelerated RI-J\cite{marij} algorithms were employed.

All subsequent calculations used previously found stable geometries and a single-point energy calculation (ridft) with a hybrid functional and a higher-quality basis set were used to obtain reliable results. For optical properties, a hybrid PBE0 XC functional~\cite{adamoReliableDensityFunctional1999, ernzerhofAssessmentPerdewBurke1999} in combination with the triple-$\zeta$-quality AO basis set with added valence and polarization functions (def2-TZVP) \cite{weigendBalancedBasisSets2005, eichkornAuxiliaryBasisSets1995, eichkornAuxiliaryBasisSets1997, weigendAccurateCoulombfittingBasis2006} was chosen. This combination of the XC functional and the basis set is known to produce an absorption spectrum of a large and diverse set of molecules with the smallest mean square error compared to many functionals, while still having acceptable computational complexity. To speed up the calculations and additionally reduce computational demand, a semi-numerical algorithm in ground (senex)\cite{https://doi.org/10.1002/jcc.22901} and excited (esenex)\cite{Holzer2020senex} state for the exchange part of the energy functional were also used alongside RI and RI-J options.

Magnetic contributions were included to the calculation of dynamic polarizabilities with a step called "mpshift". Absorption spectra were obtained by calculating the discrete electronic transitions from the ground to the 15 lowest-lying excited states, and an overall spectrum was constructed by applying a Lorentzian line-shape with 0.05 eV as FWHM to each transition and summing up all broadened line-shapes. The broadened line-shapes are summed over the energy window from \qty{1.5}{\electronvolt} to \qty{13.5}{\electronvolt}, and the spectrum is sampled at the same energies as the dynamic polarizability tensors (61 points for QM9SPol, 241 points for HOPV15).

Subsequently, electric-electric (\emph{ee}), electric-magnetic (\emph{em}), and magnetic-magnetic (\emph{mm}) static polarizabilities were calculated for all molecules. Finally, a complex \emph{ee}, \emph{em}, and \emph{mm} dynamic polarizability tensors were calculated for an equally spaced 61 frequencies from \qty{1.55}{\electronvolt} to \qty{6.2}{\electronvolt} (approximately \qty{800}{\nano\meter}-\qty{200}{\nano\meter} in wavelength). The computational costs for the different calculations are depicted in Table~\ref{tbl:si_comp_cost}. We see that the most important quantum chemistry result for our OPV simulations (the dynamic polarizability tensors) are computationally the most demanding part. Thus, to generate additional data for training the neural network, which can replace those expensive calculations, we applied the efficient AAA algorithm\cite{doi:10.1137/16M1106122, https://doi.org/10.1002/lpor.202400584, https://doi.org/10.1002/adts.202400989} allowing us to interpolate the calculated 61 frequency to a total of 241 frequency spanning the same spectral window (\qty{1.55}{\electronvolt} - \qty{6.2}{\electronvolt}). This comes at virtually no additional computational cost, whilst improving the overall training of the neural network for dynamic polarizability predictions. 

\paragraph*{HOPV15 set}

A total of 347 organic molecules with characteristic absorption in the visible (Vis) and/or ultraviolet (UV) part of the electromagnetic spectrum were extracted from the Harvard Organic Photovoltaic database (HOPV) for further studies in the context of the multi-scale multi-domain simulations of the complete OPV device stack. For all 347 molecules, the same procedure for quantum chemistry calculations was used as for the molecules contained in the QM9SPol dataset, which is described above in Subsection\ref{subsec:si_QM9SPol}. The only difference is that for obtaining the absorption spectra we calculated the 40 lowest-lying discrete electronic transitions from the ground to the excited states.

\paragraph*{Computational cost of the quantum chemistry calculations}

The quantum chemistry calculations were performed on a single computing node of a supercomputer. A node has two Intel Xeon Platinum 8368 processors with 38 CPU-cores each (base frequency 2.4GHz, max. turbo frequency 3.4 GHz) and 256 GB RAM. To minimize parallelization inefficiencies and total CPU-hours needed to produce training data, all calculations were performed either with a single core or with 2 cores to fit within the maximum given wall-time. Table~\ref{tbl:si_comp_cost} below summarizes step-wise and total CPU-hours for all quantum chemistry calculations of the photoactive layer.

\begin{table}[htbp]
\fontsize{10}{10}\selectfont
\centering
\caption{An overview of the computational cost of quantum chemistry calculations}
\label{tbl:si_comp_cost}
\begin{tabular}{llllll}
\toprule
DFT XC funct. &  & \textbf{QM9SPol} & \% of the & \textbf{HOPV15} & \% of the  \\
and basis set &  & (CPU-hours)  & total time & (CPU-hours)  & total time \\
\midrule
 & \# of molecules: & 5160 &  & 347 &   \\
\midrule
PBE & optimization: & 585.96 & 3.79 & 647.79 & 0.96  \\
def2-SVP & vib. frequencies: & 277.28 & 1.80 & 1041.83 & 1.55  \\
\midrule
 & ridft: & 209.99 & 1.36 & 496.51 & 0.74  \\
 & mpshift: & 129.36 & 0.84 & 225.99 & 0.34  \\
PBE0 & spectra 15 ex. st.: & 507.16 & 3.28 & 0.0 & 0.0  \\
def2-TZVP & spectra 40 ex. st.: & 0.0 & 0.0 & 5298.93 & 7.87  \\
 & stat. polarizability: & 148.36 & 0.96 & 339.23 & 0.50  \\
 & \textbf{dyn. polarizability:} & 13583.47 & \textbf{87.97} & 59299.35 & \textbf{88.05}  \\
\midrule
 & Total CPU-hours: & 15441.57 & 100.00 & 67349.63 & 100.00  \\
\midrule
 & CPU-hours/mol.: & 2.99 &  & 194.09 &  \\
\bottomrule
\end{tabular}
\end{table}

All outputs of the DFT and TD-DFT calculations are deposited in the NOMAD database, under the following DOI: \url{https://doi.org/10.17172/NOMAD.RG5M-AQT6}.

\begin{figure}[H]
  \centering
  \begin{subfigure}[t]{0.45\textwidth}
    \centering
    \includegraphics[width=\linewidth]{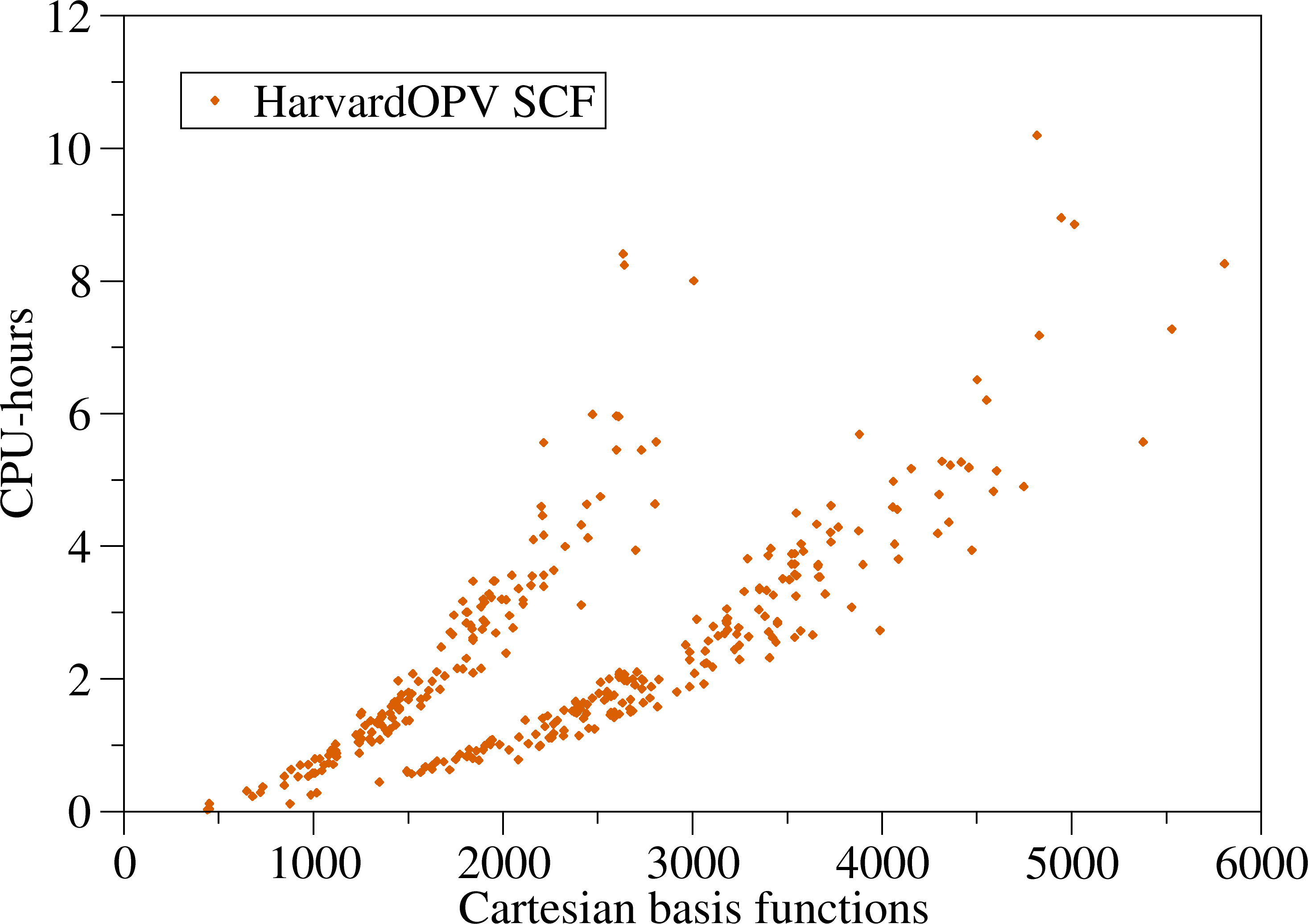}
    \caption{Computing cost of HOPV15 SCF (ridft).}
    \label{fig:si_cpu_scf}
  \end{subfigure}\hfill
  \begin{subfigure}[t]{0.45\textwidth}
    \centering
    \includegraphics[width=\linewidth]{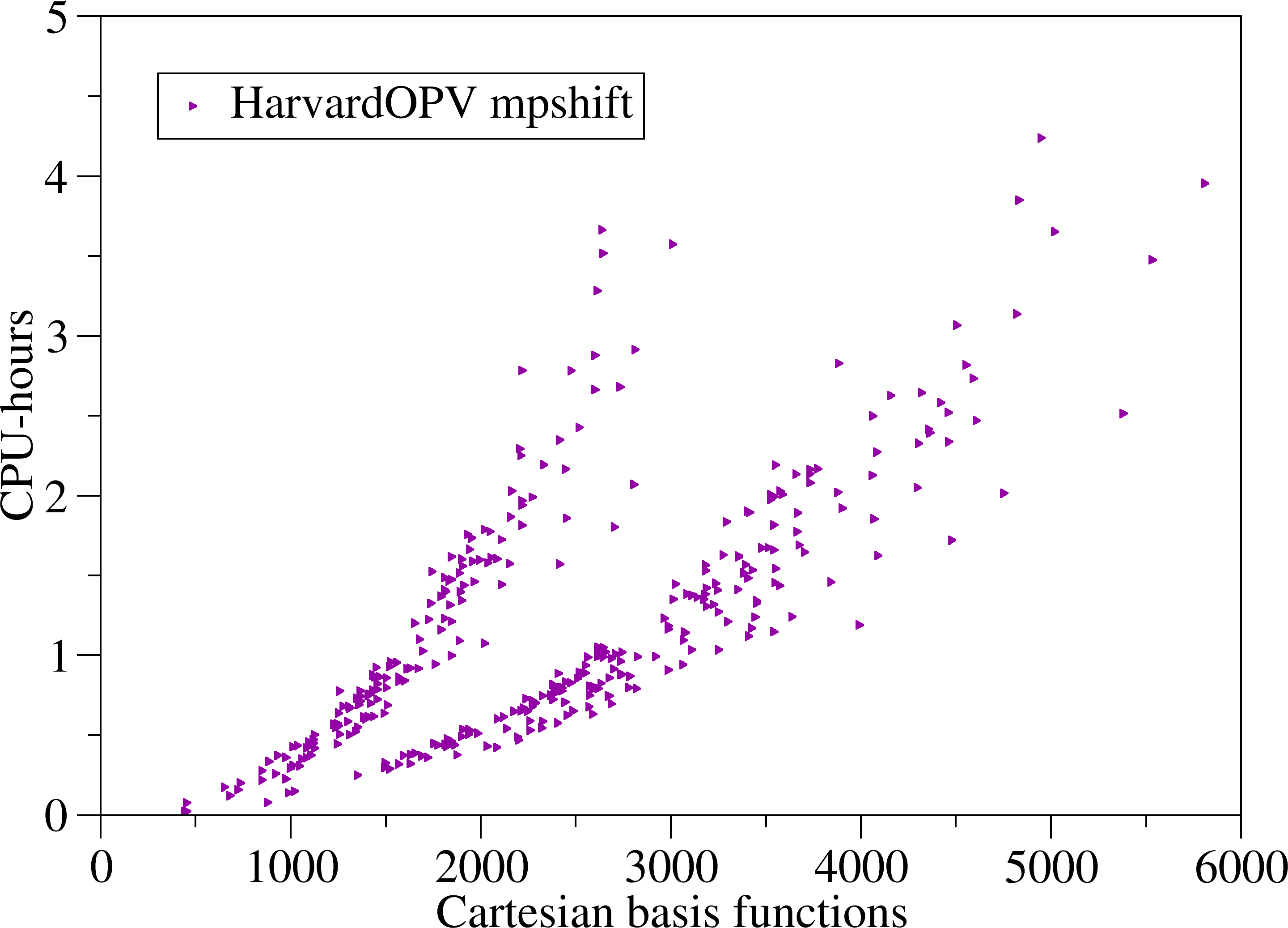}
      \caption{Computing cost of HOPV15 mpshift.}
    \label{fig:si_cpu_nmr}
  \end{subfigure}

  \vspace{0.8em}

  \begin{subfigure}[t]{0.45\textwidth}
    \centering
    \includegraphics[width=\linewidth]{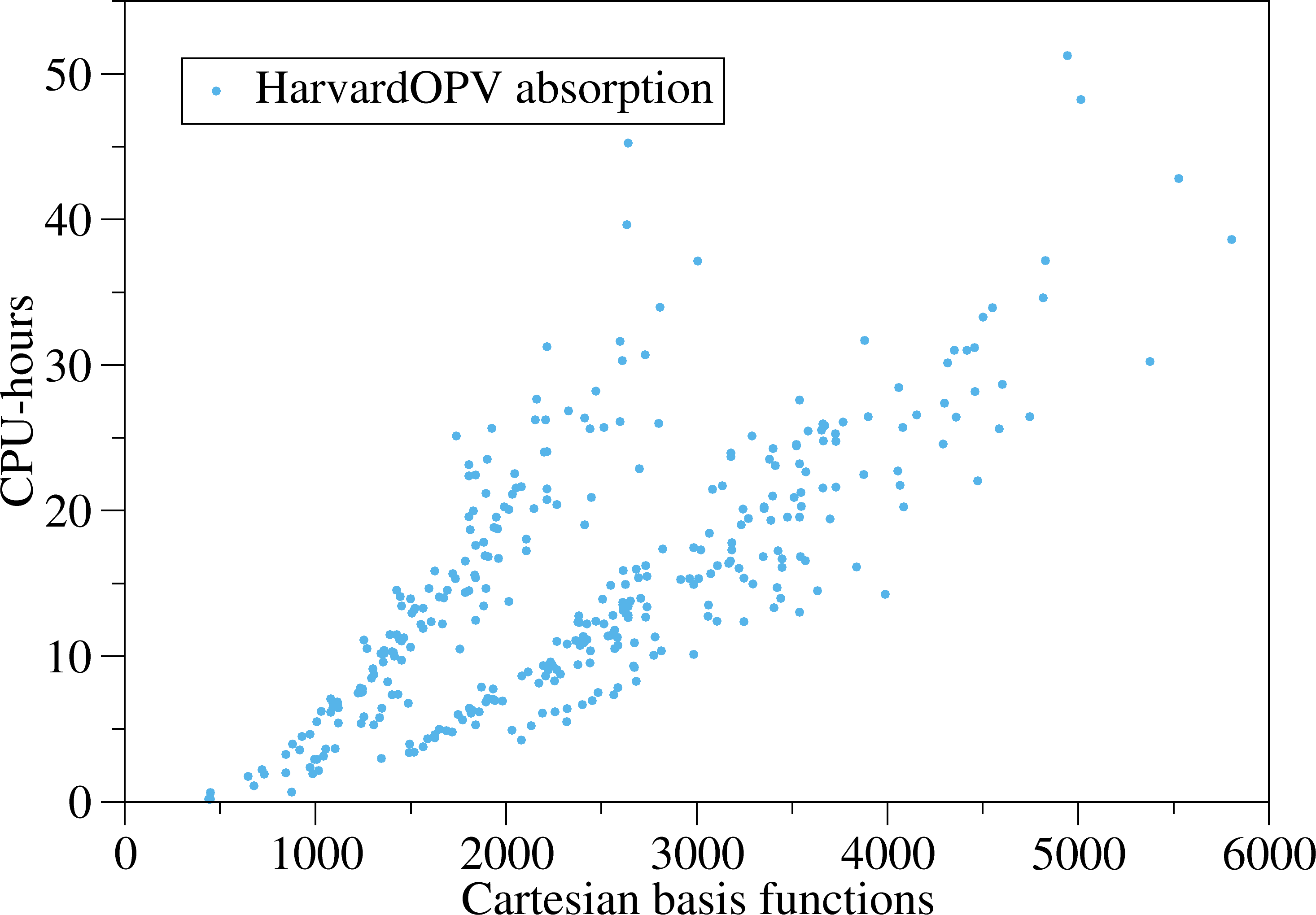}
      \caption{\centering Computing cost of HOPV15 absorption spectrum.}
\label{fig:si_cpu_abs}
  \end{subfigure}\hfill
  \begin{subfigure}[t]{0.45\textwidth}
    \centering
    \includegraphics[width=\linewidth]{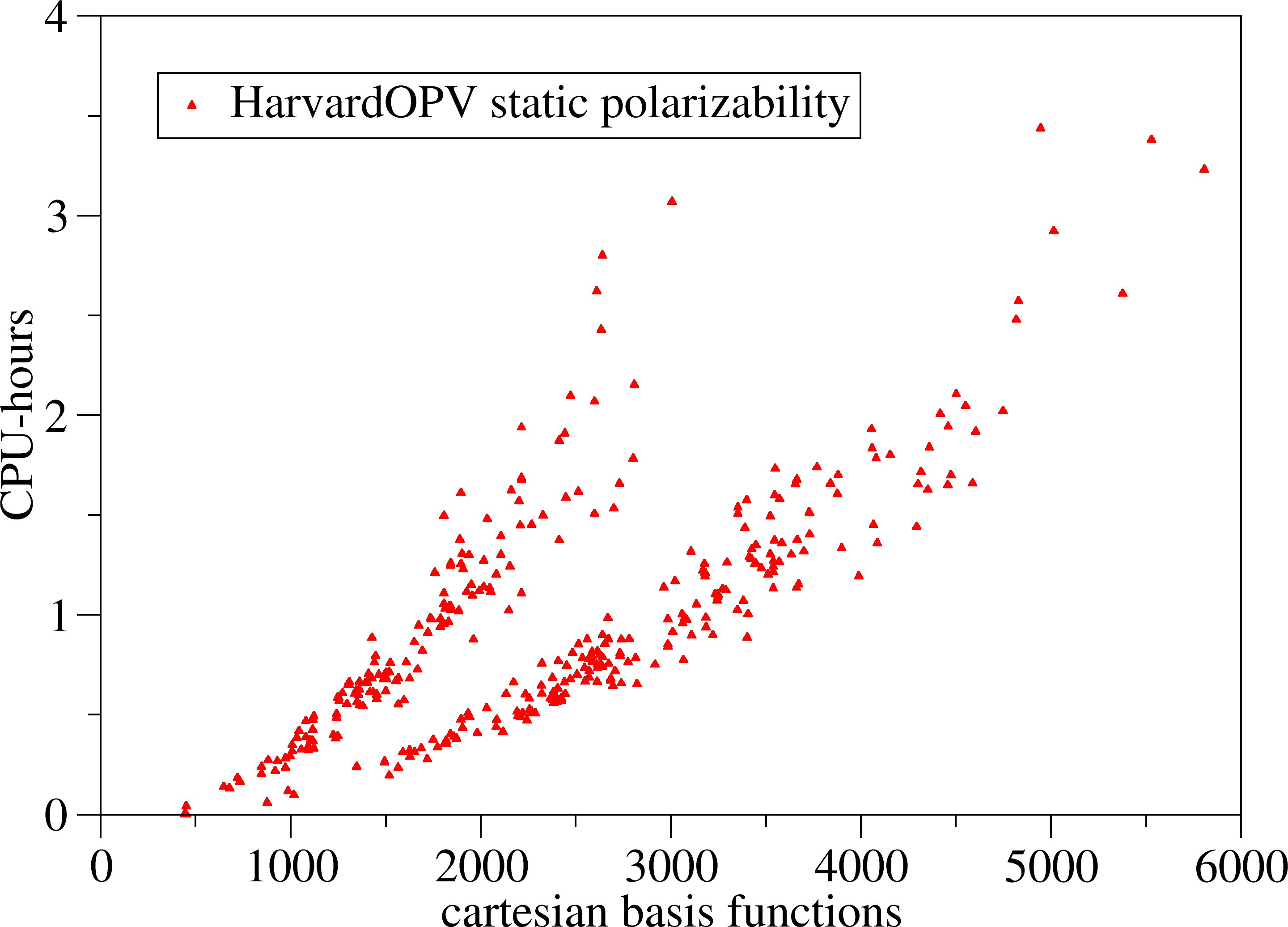}
    \caption{\centering Computing cost of HOPV15 static polarizability.}
    \label{fig:si_cpu_staticpol}
  \end{subfigure}

  \vspace{0.8em}

  \begin{subfigure}[t]{0.45\textwidth}
    \centering
    \includegraphics[width=\linewidth]{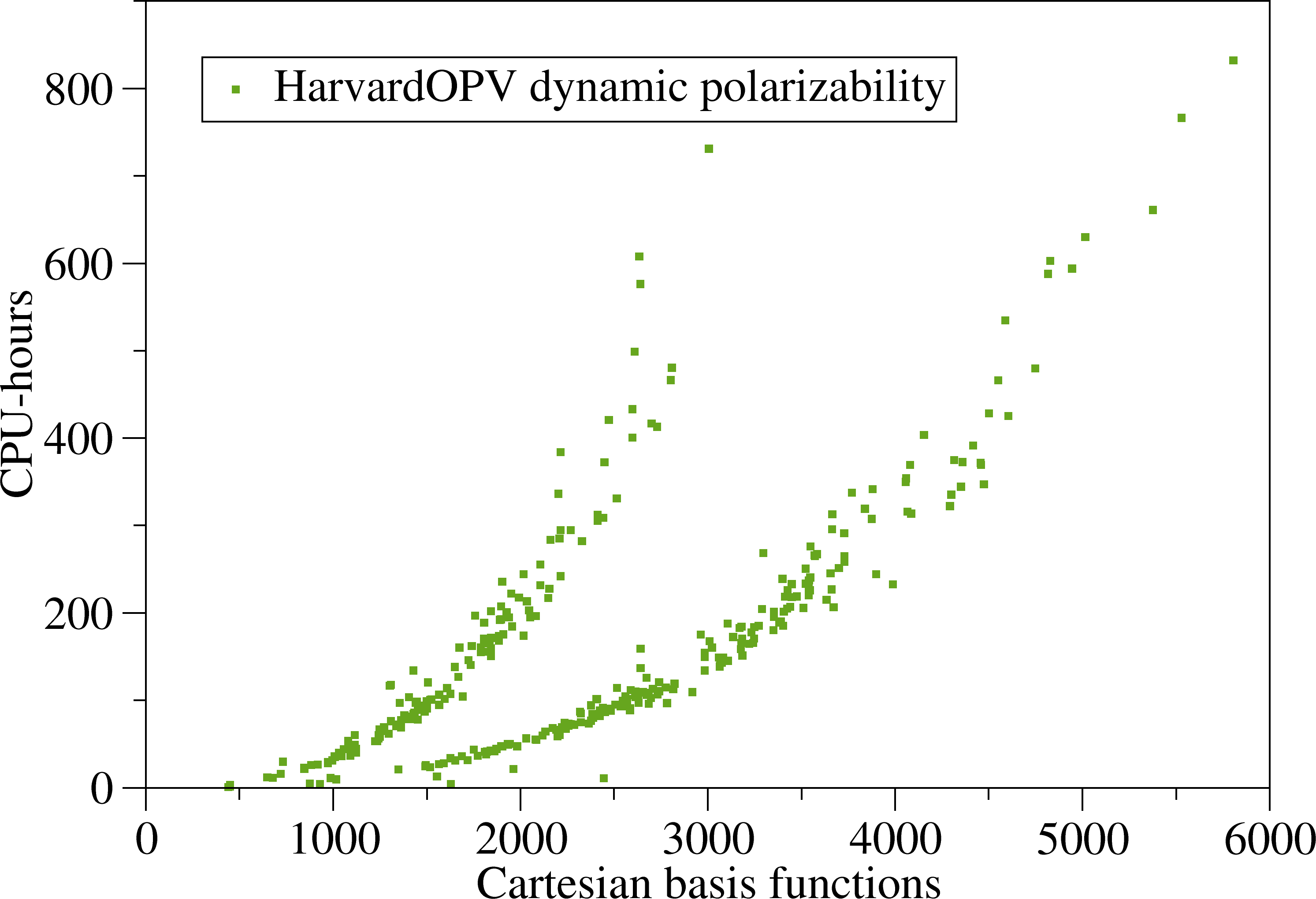}
    \caption{\centering Computing cost of HOPV15 dynamic polarizability.}
    \label{fig:si_cpu_dynpol}
  \end{subfigure}

  \caption{
    CPU time (in hours) versus number of Cartesian basis functions for all properties evaluated in the HOPV15 dataset (347 molecules).
    Each point corresponds to a single molecule.
    Among the evaluated quantities, dynamic polarizability is by far the most computationally expensive. Reproduced from~\cite{Kraus2025_Master}.
  }
  \label{fig:si_cpu_scaling_hopv}
\end{figure}

Figure~\ref{fig:si_cpu_scaling_hopv} shows the CPU time as a function of the number of Cartesian basis functions, which serves as a proxy for molecular size, for the HOPV15 dataset. The cost of computing dynamic polarizabilities clearly dominates, exceeding 800 CPU hours for the largest systems, while other properties such as static polarizability and SCF remain below 10 hours. In all five panels, two distinct bands of points are visible, hinting at two structural subsets of molecules whose electronic structure converges with systematically different effort; a closer inspection of molecular motifs is left for future work.

\subsubsection*{Details of the quantum chemistry calculations of donor polymer PM6}
To obtain the electronic ground-state structure of the PM6 polymer donor material, we considered two monomers as the molecular model. Structural molecular optimization was performed with the same hybrid PBE0 XC functional~\cite{adamoReliableDensityFunctional1999, ernzerhofAssessmentPerdewBurke1999} and the triple-$\zeta$-quality basis set with additional valence and polarization functions (def2-TZVP) \cite{weigendBalancedBasisSets2005, eichkornAuxiliaryBasisSets1995, eichkornAuxiliaryBasisSets1997, weigendAccurateCoulombfittingBasis2006} as above. Subsequent calculations of all optical properties, like the dynamic polarizabilities, used a coulomb-attenuated long-range-corrected hybrid cam-B3LYP functional~\cite{YANAI200451} with the def2-TZVP basis set. This choice ensures a correct description of long-range effects and charge-transfer states of the polymer material.

For PM6, the dynamic polarizabilities were evaluated in a wavelength range from \qty{300}{\nano\meter} to \qty{1200}{\nano\meter}, in increments of \qty{2}{\nano\meter}. This is the equivalent of \qty{1.033}{\electronvolt} to \qty{4.133}{\electronvolt}. In addition, we introduce a Lorentzian line-shape broadening factor of \qty{0.1}{\electronvolt}. We construct the T-matrices from the dynamic polarizability tensors, as described below in the \emph{Dynamic polarizability to T-matrix} section.

\subsubsection*{Molecular model of PM6 and models of HOPV molecules}

To understand the optical response of the photoactive bulk material, the optical responses on the molecular level are investigated. Since the morphological complexity of the photoactive material is high, we assume a simplified model. We treat the donor and the absorber molecules separately to calculated their optical properties, rather than directly modeling them as a mixture on the molecular level. This means that for the PM6:HOPV system, we define the structural properties of the molecular models separately and use those as the basis for all our further calculation steps. In a later step, we then mix the materials on a macroscopic level. 

\begin{figure}[htbp]
  \centering
  
      {\includegraphics[width=0.7\textwidth]{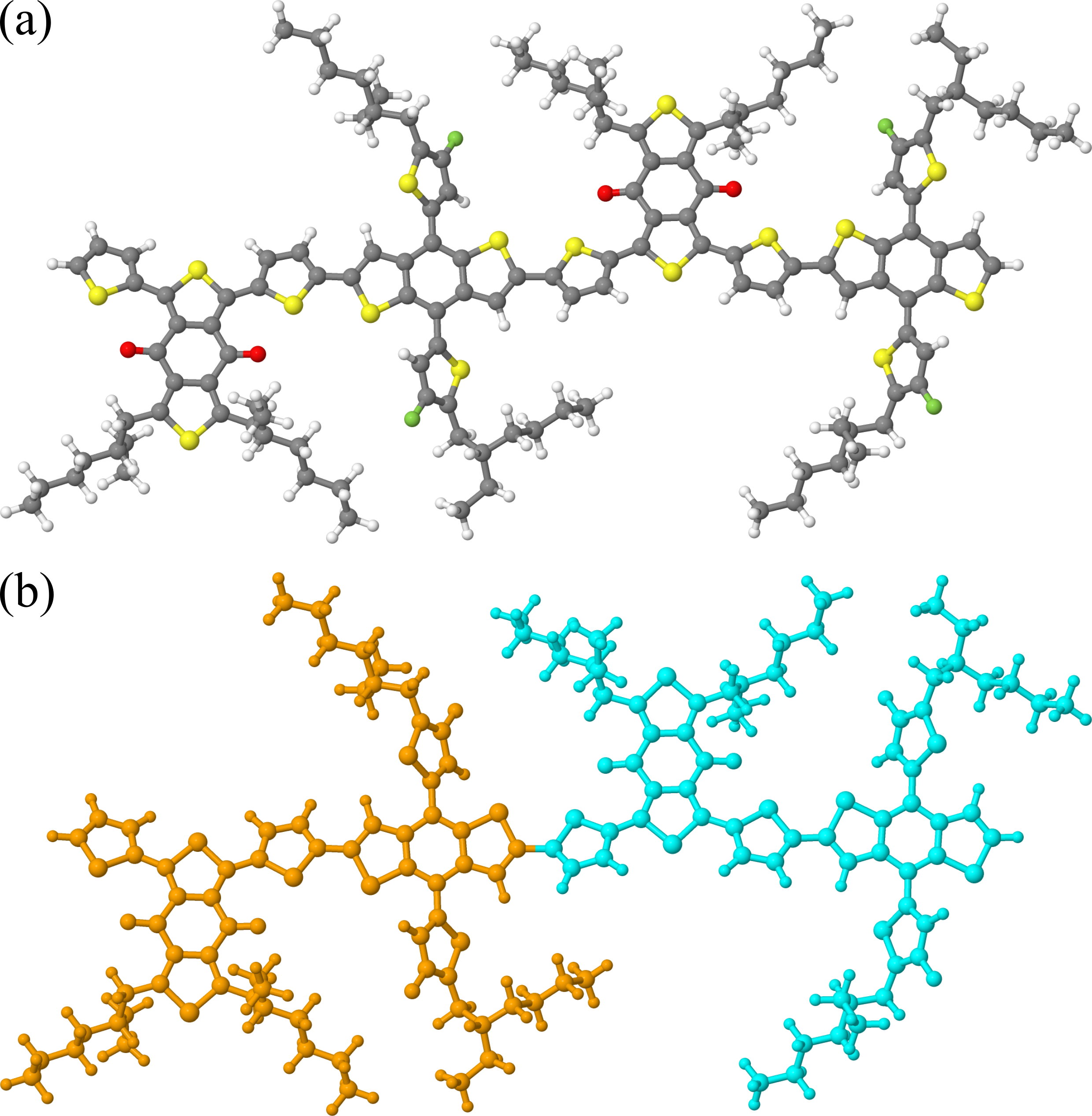}}%
  
  \hfill
  \caption
    {\textbf{(a)} This subfigure shows the DFT optimized PM6 polymer model structure. White, gray, blue, red, yellow, and green balls represent H, C, N, O, S, and F atoms. \textbf{(b)} Again, this subfigure shows the PM6 polymer structure. This time, the two individual monomers are color-marked to visualize the two monomers. Reproduced from Ref.~\cite{Schubert2025-digitaltwin}.}
  \label{fig:si_PM6_model_structure}
\end{figure}

The donor polymer PM6 is short for Poly[(2,6-(4,8-bis(5-(2-ethylhexyl-3-fluoro)thiophen-2-yl)-benzo[1,2-b:4,5-b’]dithiophene))-alt-(5,5-(1’,3’-di-2-thienyl-5’,7’-bis(2-ethylhexyl)benzo[1’,2’-c:4’,5’-c’]dithiophene-4,8-dione)]. We build a molecular representation of the polymer, consisting of two identical monomer molecules. The second monomer is connected to the first with a covalent C-C bond and rotated by \qty{180}{\degree} around the polymer chain. Overall, this structure has 314 atoms and has lattice constants of \qty{3.98}{\nano\meter}, \qty{0.44}{\nano\meter}, and \qty{2.28}{\nano\meter}. The final, geometry optimized structure is visualized in Figure~\ref{fig:si_PM6_model_structure}(a). For clarity, Figure~\ref{fig:si_PM6_model_structure}(b) shows the individual monomer units highlighted in different colors.

For each HOPV molecule, the exact atoms and their positions are known. With the help of the cavity volume constructed by the Gaussian Charge Model COSMO-GCM implicit solvation, we obtain a rough estimate of the volume of the single molecule in a dense packing within the possible crystal lattice. Thanks to reference molecules, where experimental values of volumes are known from the CIF database, we can determine an estimator for the correction factor of the calculated volume of the single molecule from COSMO-GCM. This factor is calculated to be 9.6\% on average. The data used to find the correction factor is presented in the Table~\ref{tbl:si_comp_density_correction}. We observe that the factor still does not account for the total lattice, and a significant overlap of the unit cells might be present during the homogenization procedure due to molecules having one spatial extension much larger than the other dimension. This can yield nonphysical effective parameters due to the significant overlap of the Rayleigh spheres describing the molecule in scattering theory. Therefore, to mitigate such possibilities, we enlarged the lattice constants by an additional factor of $1.7$. 

\begin{landscape}
\begin{table}[p]
\fontsize{10}{10}\selectfont
\centering
\caption{Volume correction of HOPV15 molecules for density prediction based on COSMO-GCM.}
\label{tbl:si_comp_density_correction}
\begin{tabular}{llllllllllll}
\toprule
Mol. & CIF mass & calc. mass & CIF vol. & CIF dens. & COSMO vol. & COSMO dens. & ratio & COSMO vol. corr. & corr. dens. & corr. ratio  \\
         & kg/mol. & kg/mol. & \AA$^3$/mol. & g/cm$^3$ &  \AA$^3$/mol. & g/cm$^3$ & COSMO/CIF & 1.096 & g/cm$^3$ &   \\
\midrule
54 & 5.652E-25 & 5.652E-25 & 424.4 & 1.332 & 394.07 & 1.434 & 1.077 & 431.90 & 1.309 & 0.983  \\
131 & 7.250E-25 & 7.249E-25 & 454.925 & 1.594 & 443.22 & 1.636 & 1.026 & 485.77 & 1.492 & 0.936  \\
199 & 5.454E-25 & 5.454E-25 & 373.375 & 1.461 & 349.87 & 1.559 & 1.067 & 383.46 & 1.422 & 0.974  \\
\midrule
160$^*$ & 1.417E-24 & 1.417E-24 & 1015.95 & 1.395 & 880.04 & 1.610 & 1.154 & 964.53 & 1.469 & 1.053  \\
161$^*$ & 8.574E-25 & 8.573E-25 & 472.15 & 1.816 & 428.85 & 1.999 & 1.101 & 470.02 & 1.824 & 1.004  \\
220$^*$ & 1.862E-25 & 1.862E-25 & 126.6525 & 1.470 & 123.99 & 1.502 & 1.021 & 135.89 & 1.370 & 0.932  \\
243$^*$ & 8.521E-25 & 8.520E-25 & 416.975 & 2.043 & 381.31 & 2.234 & 1.093 & 417.92 & 2.039 & 0.998  \\
245$^*$ & 5.630E-25 & 5.629E-25 & 290 & 1.941 & 266.50 & 2.112 & 1.088 & 292.09 & 1.927 & 0.993  \\
267$^*$ & 1.414E-24 & 1.425E-24 & 1040 & 1.359 & 932.61 & 1.528 & 1.124 & 1022.15 & 1.394 & 1.026  \\
269$^*$ & 1.561E-24 & 1.609E-24 & 1007.15 & 1.550 & 875.76 & 1.837 & 1.185 & 959.83 & 1.676 & 1.082  \\
287$^*$ & 1.104E-24 & 1.104E-24 & 746.495 & 1.479 & 671.92 & 1.643 & 1.111 & 736.43 & 1.499 & 1.014  \\
308$^*$ & 9.276E-25 & 9.275E-25 & 689.2 & 1.346 & 625.51 & 1.483 & 1.102 & 685.56 & 1.353 & 1.005  \\
\bottomrule
\end{tabular}\\
\footnotesize{$^*$ Molecules from the CIF database, which are structurally similar, but not exactly the same as the ones in the HOPV15 dataset.}\\
\end{table}
\end{landscape}

\subsubsection*{Dynamic polarizability to T-matrix}\label{sec:si_dynamic_pol_to_tmatrix}
The transition matrix~\cite{Waterman_Tmatrix} describes light-matter interaction of a single object~\cite{Benedikt_multiscale, Benedikt_homogenization}. The T-matrix is helpful to transition our knowledge from the microscopic, molecular level to the macroscopic, bulk material level. The T-matrix can be defined in the helicity basis as 

\begin{align}\label{eq:si_PolToTMat}
\mathbf{T} = \begin{pmatrix}
\mathbf{T}_{\mathrm{NN}}^{}&\mathbf{T}_{\mathrm{NM}}^{}\\
\mathbf{T}_{\mathrm{MN}}^{}&\mathbf{T}_{\mathrm{MM}}^{}
\end{pmatrix}
=&\frac{\mathrm{i}c_{\mathrm{h}}^{}Z_{\mathrm{h}}^{}k_{\mathrm{h}}^3}{6\pi}
\begin{pmatrix}
\mathbf{C}\left(\bm{\alpha}_{\mathrm{ee}}^{}\right)\mathbf{C}^{-1}&\mathbf{C}\left(-\mathrm{i}\bm{\alpha}_{\mathrm{em}}^{}/Z_{\mathrm{h}}^{}\right)\mathbf{C}^{-1}\\
\mathbf{C}\left(\mathrm{i}\bm{\alpha}_{\mathrm{me}}^{}/c_{\mathrm{h}}^{}\right)\mathbf{C}^{-1}&\mathbf{C}\left(\bm{\alpha}_{\mathrm{mm}}^{}/(c_{\mathrm{h}}^{}Z_{\mathrm{h}}^{})\right)\mathbf{C}^{-1}
\end{pmatrix},
\end{align}

with $\mathbf{C}$ a 3x3 matrix converting tensors from the Cartesian basis to the spherical basis, $c_{\mathrm{h}}^{}=1/\sqrt{\varepsilon_{\mathrm{h}}^{}\mu_{\mathrm{h}}^{}}$ the speed of light in the surrounding medium, and $Z_{\mathrm{h}}^{}=\sqrt{\mu_{\mathrm{h}}^{}/\varepsilon_{\mathrm{h}}^{}}$ the wave impedance in the surrounding medium. $\bm{\alpha}_{\nu \nu'}$ are the 3x3 dipolar polarizability tensors, where subscript \emph{ee} denotes the electric-electric tensor, \emph{em} the electric-magnetic tensor, \emph{me} the magnetic-electric tensor, and \emph{mm} the magnetic-magnetic tensor~\cite{Fernandez-Corbaton:2020}. Among other quantum chemistry methods, the polarizability tensors  $\bm{\alpha}_{\nu \nu'}$ of the molecular materials can be calculated with TD-DFT. They are frequency-dependent and are presented in Cartesian coordinates. 

\begin{figure}[H]
  \centering
  \includegraphics[width=0.30\textwidth]{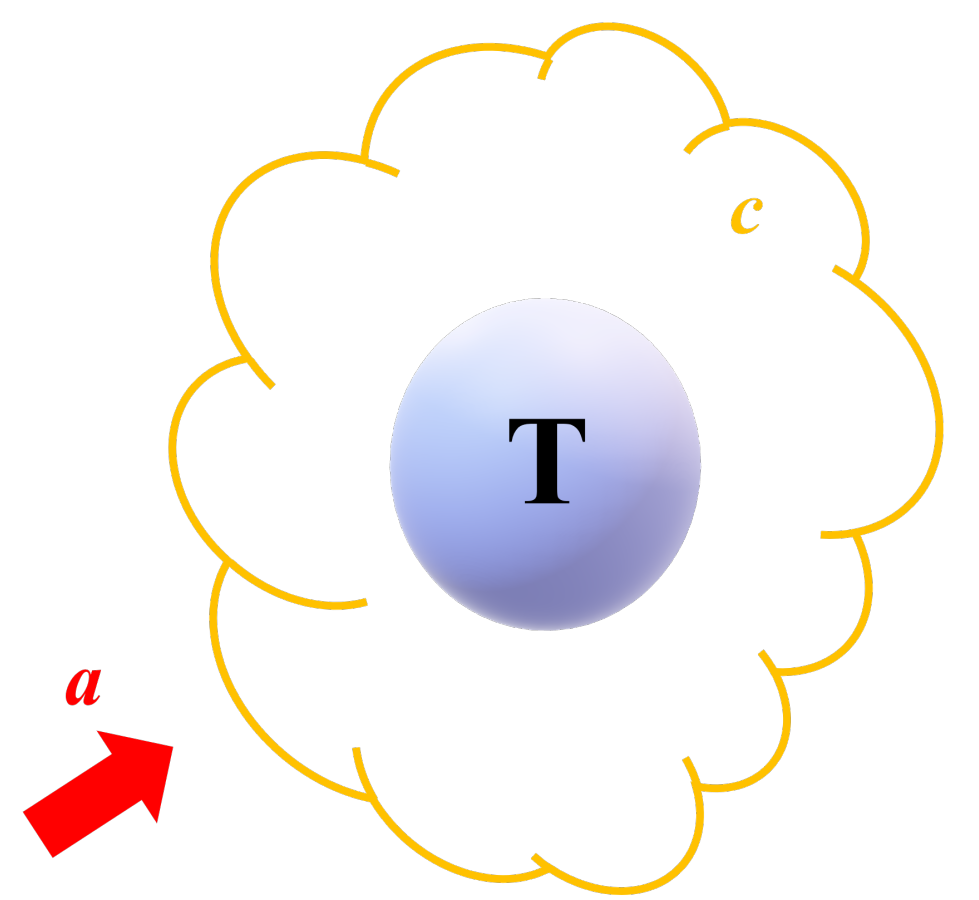}
  \caption{Illustration of the scattering process described with the
           $T$-matrix~\cite{Waterman_Tmatrix}: an incoming plane wave is expanded into vector-spherical waves with coefficients $a$, interacts with the object represented by $\mathbf{T}$, and emerges as a scattered field with coefficients $c$ satisfying $\mathbf{c} = \mathbf{T}\,\mathbf{a}$. Reproduced from~\cite{Zerulla2024_Diss}.}
  \label{fig:si_Tmatrix_schematic}
\end{figure}

Once we construct the T-matrix of a molecule, we can use it to embed the molecule in a photonic device of which we want to calculate the optical response with Maxwell simulations. 

The calculated and predicted T-matrices from quantum chemistry simulations and the neural network for the optical simulations can be provided on a reasonable request.

\subsubsection*{Homogenization}
The homogenization approach, which we used here, is based on the effective T-matrix (\textrm{\textbf{T}$_{\textrm{eff}}$}). The T-matrix \textrm{\textbf{T}$_{\textrm{molecule}}$} of the individual molecular units is used to construct the collective response of the bulk materials. Under the assumption of plane wave illumination, \textrm{\textbf{T}$_{\textrm{eff}}$} consolidates non-local effects into an effectively local description. It is defined as

\begin{equation}\label{eq:si_T_eff}
    \textrm{\textbf{T}$_{\textrm{eff}}(\textbf{k})$} =(I - \textrm{\textbf{T}$_{\textrm{molecule}}$} \sum_{\textbf{R}\neq0} 
    \textbf{C}^{(3)}(-\textbf{R})\mathrm{e}^{i\textbf{k}\cdot\textbf{R}})^{-1} \textrm{\textbf{T}$_{\textrm{molecule}}$}
\end{equation}
with \textrm{\textbf{T}$_{\textrm{molecule}}$} the T-matrix of the considered molecule, placed at lattice sites \textbf{R}, with \textbf{k} the total wave vector and $\textbf{C}^{(3)}(-\textbf{R})$ matrices the electromagnetic coupling between the origin and the lattice sites \textbf{R}~\cite{Benedikt_homogenization}. The lattice sum in equation~\ref{eq:si_T_eff} is computed with the help of the \emph{treams} framework \cite{treams_Beutel}.
The effective material parameters, such as the permittivity \textrm{$\bm{\varepsilon}_{\textrm{eff}}$} and the permeability \textrm{$\bm{\mu}_{\textrm{eff}}$}, can be derived from the dipolar part of \textrm{\textbf{T}$_{\textrm{eff}}$}~\cite{Benedikt_homogenization}. The reason behind this is the bijective link between the linear 6 × 6 local bi-anisotropic model, relating electric displacement \textbf{D} and magnetic flux density \textbf{B} to the electric field \textbf{E} and magnetic field \textbf{H} in the frequency domain, and the dipolar components of \textrm{\textbf{T}$_{\textrm{eff}}$}.

We calculate \textrm{\textbf{T}$_{\textrm{eff}}$} (and thus also the effective material parameters) separately for PM6 and each HOPV molecule. The total effective permittivity of the photoactive material blend is calculated by weighting the effective permittivity values from PM6 and the HOPV molecule in a volume ratio of 1:1. 

\subsubsection*{Permittivity values PM6} 
\begin{figure}
    \centering
    \includegraphics[width=0.75\linewidth]{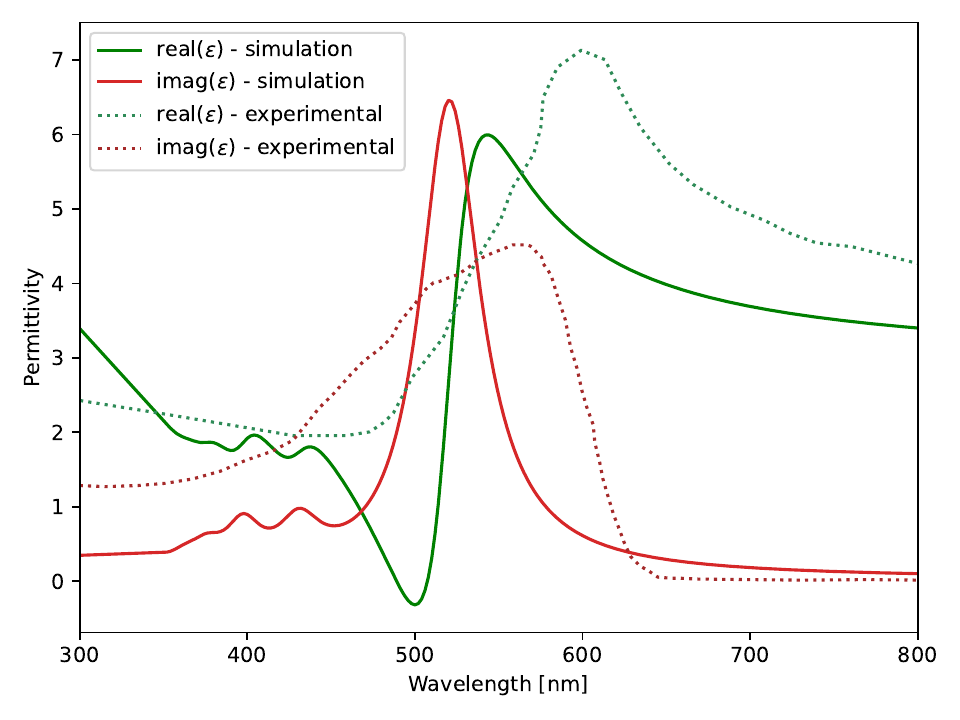}
    \caption{Permittivity values \textrm{$\bm{\varepsilon}$} for PM6. Solid lines computed with our computational approach are named \emph{simulation}. The simulation data are reproduced from Ref.~\cite{Schubert2025-digitaltwin}. The dotted lines are measured data for PM6, taken from Figure~4 of Ref.~\cite{PM6_experimental_data_2023}, and are named \emph{experimental}.}
    \label{fig:si_PM6_comparison}
\end{figure}

All the above steps were performed for the molecule PM6 as indicated in the sections before. Figure~\ref{fig:si_PM6_comparison} shows the real and imaginary parts of the permittivity values \textrm{$\bm{\varepsilon}$} of PM6, which we calculated with our computational, TD-DFT-based methods. In addition, we plot corresponding experimental values for comparison. They were measured by Ref.~\cite{PM6_experimental_data_2023} (see \emph{experimental} in  Figure~\ref{fig:si_PM6_comparison}). We observe that the placement of the peaks on the wavelength scale matches between simulation and experiment, especially for the imaginary permittivity part. Furthermore, the order of magnitude of the peaks matches well between computed and measured values. We use the simulated permittivity values \textrm{$\bm{\varepsilon}$} of PM6 for all further calculations. 

\subsubsection*{Stack architecture}
\begin{figure}
    \centering
    \includegraphics[width=0.13\textwidth]{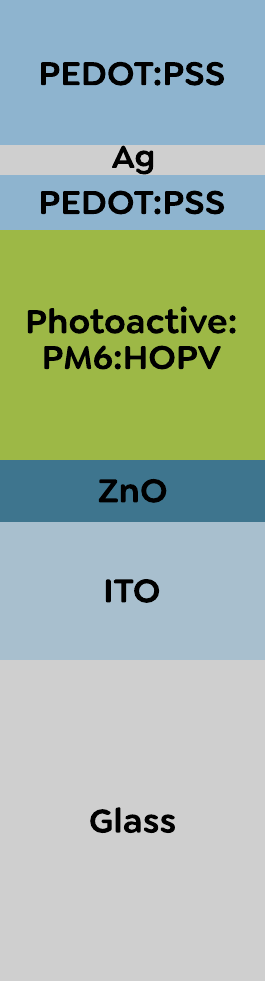}
    \caption{Device architecture of the considered organic solar cell. The stack thicknesses are not drawn to scale and measure \qty{1}{\micro\meter} of glass, \qty{150}{\nano\meter} of ITO, \qty{30}{\nano\meter} of ZnO, \qty{200}{\nano\meter} of photoactive material, \qty{50}{\nano\meter} of PEDOT:PSS, a \qty{10}{\nano\meter} layer of Ag, and finally \qty{140}{\nano\meter} of PEDOT:PSS. The photoactive material is made of PM6 and a molecule from the HOPV dataset in a 1:1 blend ratio in volume. This figure is reproduced from Ref.~\cite{Schubert2025-digitaltwin}.}
    \label{fig:si_solar_cell_stack}
\end{figure}

The considered solar cell stack is shown in Figure~\ref{fig:si_solar_cell_stack}. Sunlight enters the stack through the glass substrate with a thickness of \qty{1}{\micro\meter}. It is followed by: \qty{150}{\nano\meter} ITO, \qty{30}{\nano\meter} ZnO, \qty{200}{\nano\meter} of photoactive material, \qty{50}{\nano\meter} of PEDOT:PSS, \qty{10}{\nano\meter} Ag and \qty{140}{\nano\meter} PEDOT:PSS. All layers are assumed to be stratified isotropic media. The Ag layer is an approximation for a semi-transparent back contact of silver nanowires. The photoactive material is defined as a 1:1 blend ratio in volume of PM6 and one molecule of the HOPV test dataset. PM6 thereby acts as a donor polymer and the HOPV molecule as an acceptor polymer. 

\subsubsection*{Generation rate and S-matrix calculation}
As stated in the manuscript, the generation rate $G$ is defined as 
\begin{equation} \label{eq:si_gen_rate}
G = \mathrm{e} \int_{x}
    \int_{\lambda} P_{\mathrm{abs}}(x)\, 
    f(\lambda)\, \mathrm{d}\lambda\, \mathrm{d}x,
\end{equation}

with $\mathrm{e}$ the elementary charge, $P_{\mathrm{abs}}(x)$ the spatially resolved absorption, and factor $f(\lambda)$. $f(\lambda)$ is defined as $f(\lambda) = \mathrm{Irr(\lambda)}\, \frac{P_\mathrm{in}}{A_\mathrm{in}}$, where $\mathrm{Irr}$ is the irradiance at wavelength $\lambda$ (according to the standard AM 1.5 spectrum) and $P_\mathrm{in}$ is the power input per area $A_\mathrm{in}$. Note that equation~\ref{eq:si_gen_rate} assumes the internal quantum efficiency to be 100~\%, implying that in the photoactive layer all absorbed photons are converted into free charge carriers. Furthermore, we assume normal incidence of the light for these considerations and a photoactive layer thickness of \qty{200}{\nano\meter}. We evaluate the generation rates over a wavelength range between \qty{300}{\nano\meter} to \qty{800}{\nano\meter}.

The spatially resolved absorption $P_{\mathrm{abs}}(x)$ is given by the expression
\begin{equation} \label{eq:si_P_abs}
    P_{\mathrm{abs}}(x) = -0.5\omega \left| \mathrm{E}(x)\right|^2\operatorname{Im}(\varepsilon)
\end{equation}
with $\omega$ the angular frequency $\frac{2\pi c}{\lambda}$, the field $\mathrm{E}(x)$ and the permittivity $\varepsilon$ of the material at position $x$. We can calculate the field $\mathrm{E}(x)$ via the S-matrix~\cite{S_matrix_Cotter}. Given the permittivity $\varepsilon$ and the thickness $d$ of a layer, we can analytically construct its S-matrix~\cite{S_matrix_STEFANOU199849}. The S-matrix of the full structure is then obtained through the Redheffer product~\cite{Redhaffer_star_product, Redheffer_product_original}. This enables the analysis of the photon absorption both for an individual layer of the solar cell and for the entire stack.

\subsubsection*{Some examples of HOPV molecules}

In this section, we give some examples of PM6:HOPV molecule combinations. The aim is to facilitate a better understanding of the resulting permittivity values $\varepsilon$ of the different molecules and how they influence the overall permittivity of the photoactive material. 

\begin{figure}
    \centering
    \includegraphics[width=0.99\linewidth]{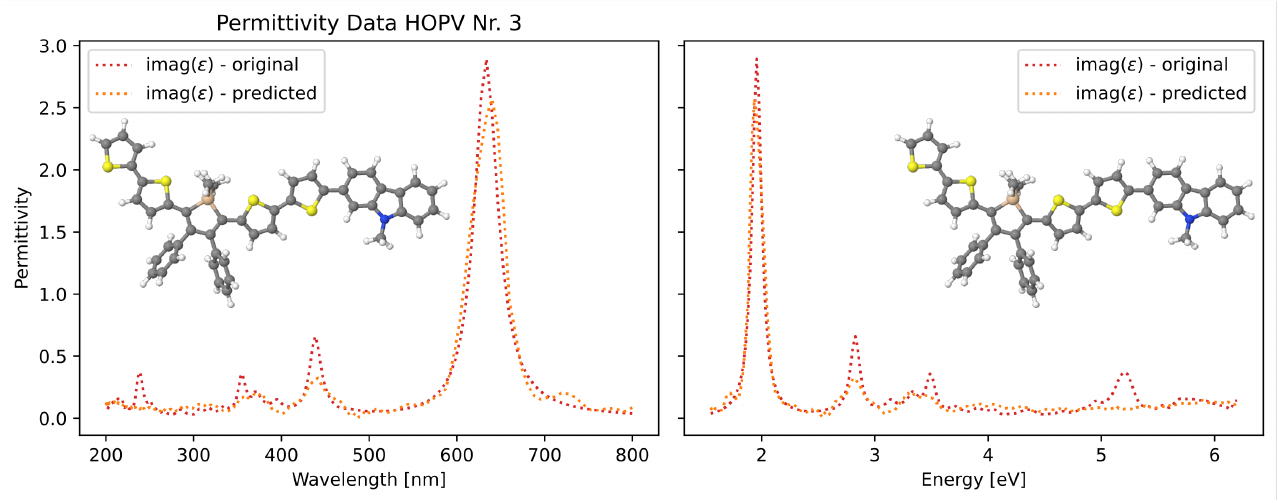}
    \caption{Imaginary part of the permittivity $\varepsilon$ of HOPV molecule number 3. The red points show the permittivity data for the \emph{original}, TD-DFT-based, case. The orange points show the imaginary part of the permittivity values for the \emph{predicted} T-matrix. We observe a clear similarity between the two cases. The left plot shows the results over the wavelength in nanometers, the right plot over the frequency in electron volts. The inserted figures show the model structure of this molecule. }
    \label{fig:si_HOPV3_alone}
\end{figure}

\paragraph*{HOPV molecule number 3}
For the HOPV molecule number 3, we observe a good agreement between the original permittivity values and the predicted ones. This is shown in Figure~\ref{fig:si_HOPV3_alone} for the imaginary parts of the permittivity values $\varepsilon$ of the molecule alone. In this context, \emph{original} means that we calculate the T-matrix via computationally expensive TD-DFT calculations. We then use this T-matrix to calculate the permittivity values $\varepsilon$ as described above. The term \emph{predicted} is used when the T-matrix is predicted by our ML-model. The predicted T-matrix is then used to find the permittivity values $\varepsilon$, according to the homogenization steps described in the previous sections. 

\begin{figure}
    \centering
    \includegraphics[width=0.7\linewidth]{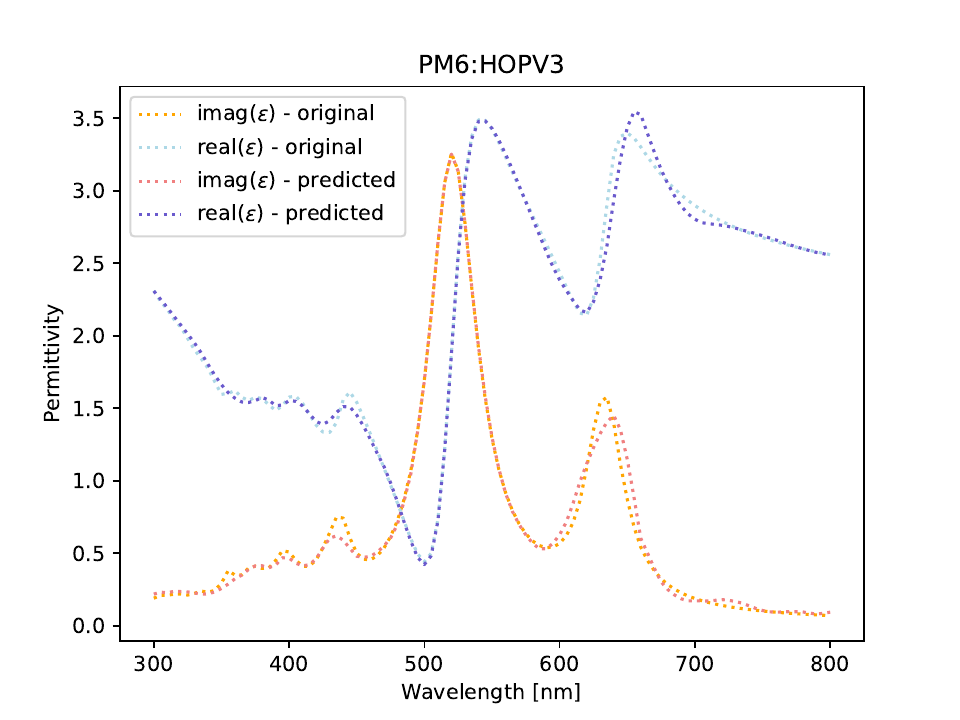}
    \caption{Permittivity values $\varepsilon$ for the photoactive bulk material PM6:HOPV3. The blue tones correspond to the real part, the red tones show the imaginary part of  $\varepsilon$. The darker color shades, red and blue, show the \emph{originally} calculated data. The lighter color shades, orange and light blue, show the results of the \emph{predictive} mode.}
    \label{fig:si_HOPV3_PM6}
\end{figure}

Next, we consider the photoactive material combination of PM6 and the HOPV molecule number~3, called PM6:HOPV3. The permittivity data for the \emph{original} case and the \emph{predicted} case are given in Figure~\ref{fig:si_HOPV3_PM6}. We recognize two distinct peaks - one corresponding to the PM6 molecule and one to the HOPV molecule number~3. 

\begin{figure}
    \centering
    \includegraphics[width=0.99\linewidth]{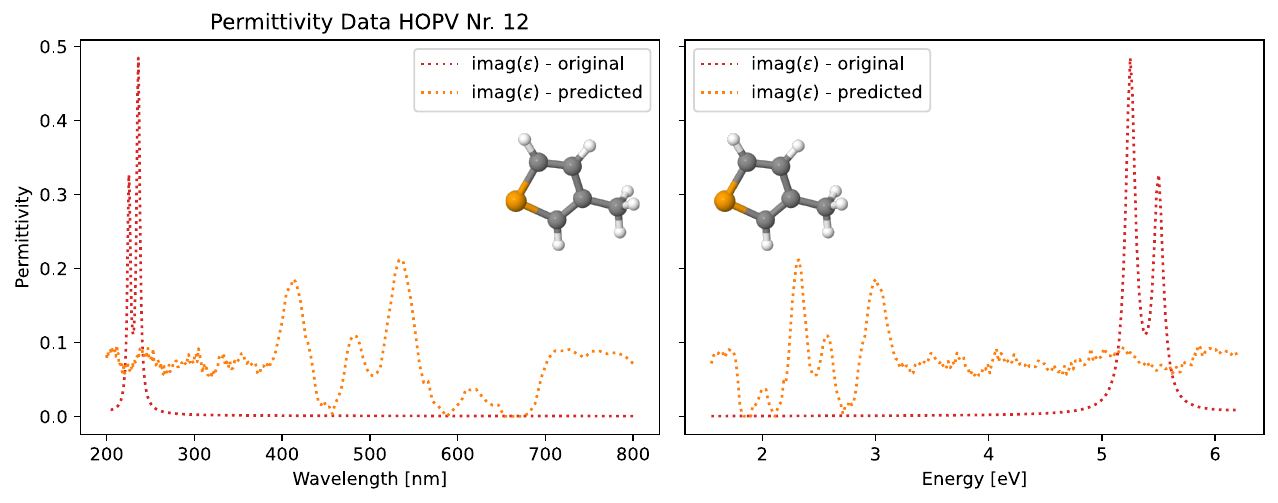}
    \caption{This figure shows the imaginary part of the permittivity $\varepsilon$ of the HOPV molecule number 12. On the left, we show the permittivity $\varepsilon$ over the wavelength. On the right, we show the permittivity $\varepsilon$ over the energy. The values stemming from the \emph{original} T-matrix calculation (in red) deviate clearly from the values from the \emph{predicted} T-matrices (in orange). Overall, the magnitude of the values is rather small. The inserted figures show the model structure of the HOPV molecule number 12. }
    \label{fig:si_HOPV12_alone}
\end{figure}

\paragraph*{HOPV molecule number 12}
Figure~\ref{fig:si_HOPV12_alone} shows the imaginary parts of the permittivity values $\varepsilon$ in the \emph{original} and the \emph{predicted} case. Again, \emph{original} refers to the case where the permittivity data stems from the TD-DFT calculated T-matrix, and \emph{predicted} to the case where the T-matrix is predicted from our model. For this example, the agreement between the original permittivity values and the predicted values is not very good. Notice, that the magnitude of the permittivity values $\varepsilon$ is very small, meaning that this specific molecule does not absorb well in the given wavelength range.

\begin{figure}
    \centering
    \includegraphics[width=0.7\linewidth]{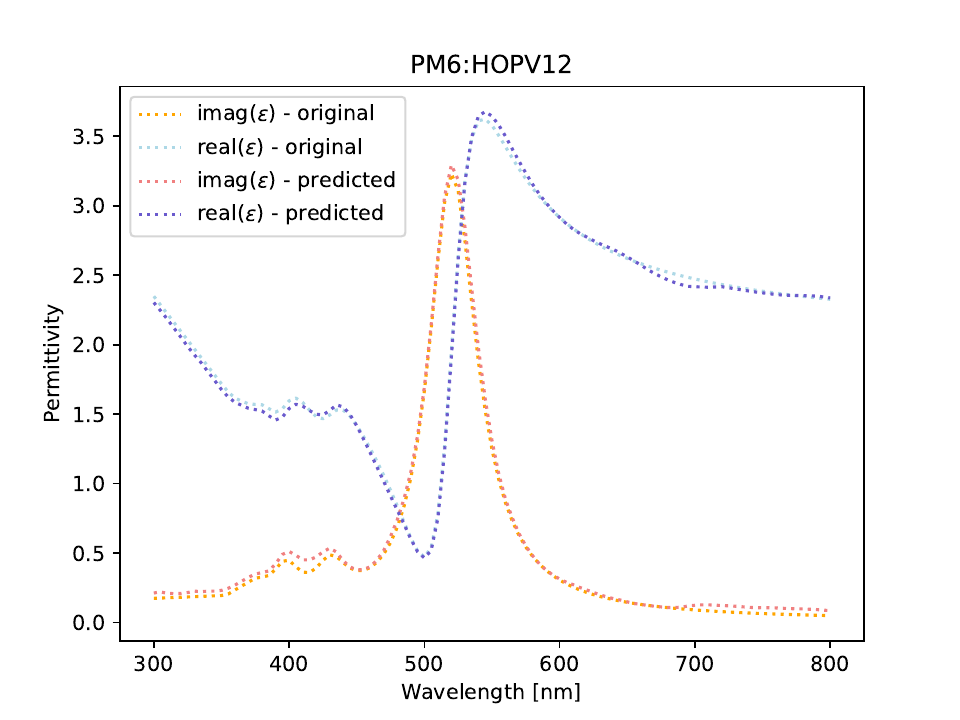}
    \caption{This figure shows the PM6:HOPV12 permittivity values $\varepsilon$. Once calculated in the \emph{original}, TD-DFT way (in darker colors) and once in the \emph{predicted} ML-based way (in lighter colors). The blue tones indicated the real parts of $\varepsilon$ and the orange tones the imaginary parts of $\varepsilon$. The plot shows that PM6 is dominant over the HOPV molecule number 12.}
    \label{fig:si_HOPV12_PM6}
\end{figure}

In addition, we look at the permittivity values $\varepsilon$ for the photoactive bulk material PM6:HOPV12, see Figure~\ref{fig:si_HOPV12_PM6}. In this case, we only observe one peak in the permittivity values $\varepsilon$. When taking Figure~\ref{fig:si_PM6_comparison} into account, we notice that this peak corresponds to PM6 only. This means the HOPV molecule number 12 does not add any additional absorption peak, neither in the original nor in the predicted case. Therefore, despite the strong mismatch for the HOPV molecule 12 alone, the photoactive material does not exhibit particularly different results for original or predicted values.  

\clearpage
\subsection*{References}
\printbibliography[heading=none]

%% file: manuscript.bib
@misc{Schubert2025-digitaltwin,
  options = {maxbibnames=2,minbibnames=2},
  note    = {M.~L.~Schubert and H.~Metni contributed equally},
  title   = {Towards a fully differentiable digital twin for solar cells},
  author  = {Schubert, Marie Louise and Metni, Houssam and Fischbach, Jan David and Zerulla, Benedikt and Krsti{\'c}, Marjan and Paetzold, Ulrich W. and Orooji, Seyedamir and Ronsin, Olivier J. J. and Ameslon, Yasin and Harting, Jens and Kirchartz, Thomas and Ravishankar, Sandheep and Dreessen, Chris and Kim, Eunchi and Sprau, Christian and Hussein, Mohamed and Colsmann, Alexander and Forberich, Karen and J{\"a}ger, Klaus and Friederich, Pascal and Rockstuhl, Carsten},
  year    = {2025},
  eprint  = {2512.02904},
  archivePrefix = {arXiv},
  primaryClass = {physics.comp-ph},
  url     = {https://arxiv.org/abs/2512.02904},
  howpublished = {arXiv preprint arXiv:2512.02904}
}

@article{Zhang2025_OrganicPhotodetectors,
  author  = {Zhang, Kai and Wu, Junjiang and Sun, Chunlong and Chung, Dae Sung and Geng, Yanhou and Ye, Long},
  title   = {The rising promise of organic photodetectors in emerging technologies},
  journal = {Nature Reviews Materials},
  year    = {2025},
  doi     = {10.1038/s41578-025-00821-2},
  pages = {821}
}

@article{friederich2017rational,
  title={Rational in silico design of an organic semiconductor with improved electron mobility},
  author={Friederich, Pascal and G{\'o}mez, Ver{\'o}nica and Sprau, Christian and Meded, Velimir and Strunk, Timo and Jenne, Michael and Magri, Andrea and Symalla, Franz and Colsmann, Alexander and Ruben, Mario and others},
  journal={Advanced Materials},
  volume={29},
  number={43},
  pages={1703505},
  year={2017},
  publisher={Wiley Online Library}
}

@article{wu2024inverse,
  title={Inverse design workflow discovers hole-transport materials tailored for perovskite solar cells},
  author={Wu, Jianchang and Torresi, Luca and Hu, ManMan and Reiser, Patrick and Zhang, Jiyun and Rocha-Ortiz, Juan S and Wang, Luyao and Xie, Zhiqiang and Zhang, Kaicheng and Park, Byung-wook and others},
  journal={Science},
  volume={386},
  number={6727},
  pages={1256--1264},
  year={2024},
  publisher={American Association for the Advancement of Science}
}

@article{Wang2024_FlexibleOrganicOptoelectronics,
  author  = {Wang, Yuanhe and Wen, Yanlong and Zhuang, Xiaoxiao and Liu, Shihao and Zhang, Letian and Xie, Wenfa},
  title   = {Flexible organic optoelectronic devices: Design, fabrication, and applications},
  journal = {APL Photonics},
  volume  = {9},
  number  = {9},
  pages   = {090901},
  year    = {2024},
  doi     = {10.1063/5.0220555}
}

@article{Nowsherwan2024_OrganicOptoelectronics,
  author  = {Nowsherwan, Ghazi Aman and Ali, Qasim and Ali, Umar Farooq and Ahmad, Muhammad and Khan, Mohsin and Hussain, Syed Sajjad},
  title   = {Advances in Organic Materials for Next-Generation Optoelectronics: Potential and Challenges},
  journal = {Organics},
  volume  = {5},
  number  = {4},
  pages   = {520-560},
  year    = {2024},
  doi     = {10.3390/org5040028}
}

@article{Ostroverkhova2016_OrganicOptoelectronics,
  author  = {Ostroverkhova, Oksana},
  title   = {Organic Optoelectronic Materials: Mechanisms and Applications},
  journal = {Chemical Reviews},
  volume  = {116},
  number  = {22},
  pages   = {13279-13412},
  year    = {2016},
  doi     = {10.1021/acs.chemrev.6b00127}
}

@article{green2014emergence,
  title={The emergence of perovskite solar cells},
  author={Green, Martin A and Ho-Baillie, Anita and Snaith, Henry J},
  journal={Nature photonics},
  volume={8},
  number={7},
  pages={506--514},
  year={2014},
  publisher={Nature Publishing Group UK London},
  doi={https://doi.org/10.1038/nphoton.2014.134}
}

@article{ColsmanetalOPV,
author = {Colsmann, Alexander and Röhm, Holger and Sprau, Christian},
title = {Shining Light on Organic Solar Cells},
journal = {Solar RRL},
volume = {4},
number = {6},
pages = {2000015},
doi = {https://doi.org/10.1002/solr.202000015},
url = {https://onlinelibrary.wiley.com/doi/abs/10.1002/solr.202000015},
eprint = {https://onlinelibrary.wiley.com/doi/pdf/10.1002/solr.202000015},
year = {2020}
}

@ARTICLE{Musaelian2023_Allegro,
  title    = "Learning local equivariant representations for large-scale
              atomistic dynamics",
  author   = "Musaelian, Albert and Batzner, Simon and Johansson, Anders and
              Sun, Lixin and Owen, Cameron J and Kornbluth, Mordechai and
              Kozinsky, Boris",
  journal  = "Nat. Commun.",
  volume   =  14,
  number   =  1,
  pages    = "579",
  month    =  feb,
  year     =  2023,
  language = "en",
  doi = {10.1038/s41467-023-36329-y}
}

@misc{satorras2022E_n,
      title={E(n) Equivariant Graph Neural Networks}, 
      author={Victor Garcia Satorras and Emiel Hoogeboom and Max Welling},
      year={2022},
      eprint={2102.09844},
      archivePrefix={arXiv},
      primaryClass={cs.LG},
      url={https://arxiv.org/abs/2102.09844}, 
}

@ARTICLE{Zou2023-DetaNet,
  title     = "A deep learning model for predicting selected organic molecular
               spectra",
  author    = "Zou, Zihan and Zhang, Yujin and Liang, Lijun and Wei, Mingzhi
               and Leng, Jiancai and Jiang, Jun and Luo, Yi and Hu, Wei",
  journal   = "Nat. Comput. Sci.",
  publisher = "Springer Science and Business Media LLC",
  volume    =  3,
  number    =  11,
  pages     = "957--964",
  month     =  nov,
  year      =  2023,
  copyright = "https://www.springernature.com/gp/researchers/text-and-data-mining",
  language  = "en",
  doi = {10.1038/s43588-023-00550-y}
}

@article{Schutt2018_SchNet,
    author = {Schütt, K. T. and Sauceda, H. E. and Kindermans, P.-J. and Tkatchenko, A. and Müller, K.-R.},
    title = "{SchNet – A deep learning architecture for molecules and materials}",
    journal = {The Journal of Chemical Physics},
    volume = {148},
    number = {24},
    pages = {241722},
    year = {2018},
    month = {03},
    issn = {0021-9606},
    doi = {10.1063/1.5019779},
    url = {https://doi.org/10.1063/1.5019779},
}

@article{Klicpera2020_DimeNet,
  author       = {Johannes Klicpera and
                  Janek Gro{\ss} and
                  Stephan G{\"{u}}nnemann},
  title        = {Directional Message Passing for Molecular Graphs},
  journal      = {CoRR},
  volume       = {abs/2003.03123},
  year         = {2020},
  url          = {https://arxiv.org/abs/2003.03123},
  eprinttype    = {arXiv},
  eprint       = {2003.03123},
  bibsource    = {dblp computer science bibliography, https://dblp.org}
}

@article{Kristof2021_PaiNN,
  author       = {Kristof T. Sch{\"{u}}tt and
                  Oliver T. Unke and
                  Michael Gastegger},
  title        = {Equivariant message passing for the prediction of tensorial properties
                  and molecular spectra},
  journal      = {CoRR},
  volume       = {abs/2102.03150},
  year         = {2021},
  url          = {https://arxiv.org/abs/2102.03150},
  eprinttype    = {arXiv},
  eprint       = {2102.03150},
  bibsource    = {dblp computer science bibliography, https://dblp.org}
}

@ARTICLE{Batzner2022-NequIP,
  title     = "E(3)-equivariant graph neural networks for data-efficient and
               accurate interatomic potentials",
  author    = "Batzner, Simon and Musaelian, Albert and Sun, Lixin and Geiger,
               Mario and Mailoa, Jonathan P and Kornbluth, Mordechai and
               Molinari, Nicola and Smidt, Tess E and Kozinsky, Boris",
  journal   = "Nat. Commun.",
  publisher = "Springer Science and Business Media LLC",
  volume    =  13,
  number    =  1,
  pages     = "2453",
  month     =  may,
  year      =  2022,
  copyright = "https://creativecommons.org/licenses/by/4.0",
  language  = "en",
  doi = {10.1038/s41467-022-29939-5}
}

@misc{2023_LibreTextChemie, title={Polarizability}, url={https://chem.libretexts.org/Bookshelves/Physical_and_Theoretical_Chemistry_Textbook_Maps/Supplemental_Modules_(Physical_and_Theoretical_Chemistry)/Physical_Properties_of_Matter/Atomic_and_Molecular_Properties/Intermolecular_Forces/Specific_Interactions/Polarizability}, journal={Chemistry LibreTexts}, publisher={Libretexts}, author={Larsen, Delmar and Cox, Kelly and Reussner, Dana}, year={2023}, month=feb}

@misc{Thomas2018-tensorFieldNetworks,
  title        = "Tensor field networks: Rotation- and translation-equivariant
                  neural networks for {3D} point clouds",
  author       = "Thomas, Nathaniel and Smidt, Tess and Kearnes, Steven and
                  Yang, Lusann and Li, Li and Kohlhoff, Kai and Riley, Patrick",
  year         =  2018,
  primaryClass = "cs.LG",
  eprint       = "1802.08219",
  url = {https://arxiv.org/abs/1802.08219},
  howpublished = {arXiv preprint}
}

@article{QM9ramakrishnan2014quantum,
  title={Quantum chemistry structures and properties of 134 kilo molecules},
  author={Ramakrishnan, Raghunathan and Dral, Pavlo O and Rupp, Matthias and von Lilienfeld, O Anatole},
  journal={Scientific Data},
  volume={1},
  year={2014},
  publisher={Nature Publishing Group},
  doi = {10.1038/sdata.2014.22}
}

@article{QM9,
author = {Ruddigkeit, Lars and van Deursen, Ruud and Blum, Lorenz C. and Reymond, Jean-Louis},
title = {Enumeration of 166 Billion Organic Small Molecules in the Chemical Universe Database GDB-17},
journal = {Journal of Chemical Information and Modeling},
volume = {52},
number = {11},
pages = {2864-2875},
year = {2012},
doi = {10.1021/ci300415d},
note ={PMID: 23088335},
URL = {https://doi.org/10.1021/ci300415d},
eprint = {https://doi.org/10.1021/ci300415d}
}

@ARTICLE{Lopez2016-HarvardPhotoDataset,
  title     = "The Harvard organic photovoltaic dataset",
  author    = "Lopez, Steven A and Pyzer-Knapp, Edward O and Simm, Gregor N and
               Lutzow, Trevor and Li, Kewei and Seress, Laszlo R and Hachmann,
               Johannes and Aspuru-Guzik, Al{\'a}n",
  journal   = "Sci. Data",
  publisher = "Springer Science and Business Media LLC",
  volume    =  3,
  number    =  1,
  pages     = "160086",
  month     =  sep,
  year      =  2016,
  language  = "en",
  doi = {10.1038/sdata.2016.86}
}

@inproceedings{Gilmer2017,
author = {Gilmer, Justin and Schoenholz, Samuel S. and Riley, Patrick F. and Vinyals, Oriol and Dahl, George E.},
title = {Neural message passing for Quantum chemistry},
year = {2017},
publisher = {JMLR.org},
booktitle = {Proceedings of the 34th International Conference on Machine Learning - Volume 70},
pages = {1263–1272},
numpages = {10},
location = {Sydney, NSW, Australia},
series = {ICML'17},
  url = {https://proceedings.mlr.press/v70/gilmer17a.html}
}

@misc{fuchs2020se3transformers3drototranslationequivariant,
      title={SE(3)-Transformers: 3D Roto-Translation Equivariant Attention Networks}, 
      author={Fabian B. Fuchs and Daniel E. Worrall and Volker Fischer and Max Welling},
      year={2020},
      eprint={2006.10503},
      archivePrefix={arXiv},
      primaryClass={cs.LG},
      url={https://arxiv.org/abs/2006.10503}, 
}

@misc{geiger2022e3nneuclideanneuralnetworks,
      title={e3nn: Euclidean Neural Networks}, 
      author={Mario Geiger and Tess Smidt},
      year={2022},
      eprint={2207.09453},
      archivePrefix={arXiv},
      primaryClass={cs.LG},
      url={https://arxiv.org/abs/2207.09453}, 
}

@ARTICLE{Falletta2025-jf,
  title    = "Unified differentiable learning of electric response",
  author   = "Falletta, Stefano and Cepellotti, Andrea and Johansson, Anders
              and Tan, Chuin Wei and Descoteaux, Marc L and Musaelian, Albert
              and Owen, Cameron J and Kozinsky, Boris",
  journal  = "Nat. Commun.",
  volume   =  16,
  number   =  1,
  pages    = "4031",
  month    =  apr,
  year     =  2025,
  language = "en",
  doi = {10.1038/s41467-025-59304-1}
}

@misc{filling2025direct,
  title   = {Direct Molecular Polarizability Prediction with {$SO(3)$} Equivariant Local Frame {GNNs}},
  author  = {Filling, Jean Philip and Post, Felix and Wand, Michael and Andrienko, Denis},
  year    = {2025},
  eprint  = {2511.07087},
  archivePrefix = {arXiv},
  primaryClass = {cs.LG},
  url     = {https://arxiv.org/abs/2511.07087},
  howpublished = {arXiv preprint arXiv:2511.07087}
}

@ARTICLE{Wang2024-zn,
  title     = "{E(n)-Equivariant} cartesian tensor message passing interatomic
               potential",
  author    = "Wang, Junjie and Wang, Yong and Zhang, Haoting and Yang, Ziyang
               and Liang, Zhixin and Shi, Jiuyang and Wang, Hui-Tian and Xing,
               Dingyu and Sun, Jian",
  journal   = "Nat. Commun.",
  publisher = "Springer Science and Business Media LLC",
  volume    =  15,
  number    =  1,
  pages     = "7607",
  month     =  sep,
  year      =  2024,
  copyright = "https://creativecommons.org/licenses/by-nc-nd/4.0",
  language  = "en",
  doi = {10.1038/s41467-024-51886-6}
}

@phdthesis{Zerulla2024_Diss,
  author  = "Benedikt Alexander Raphael Zerulla",
  title   = "Photonic Devices Made from Molecular Materials: A Multi-Scale Modeling Approach",
  school  = "Karlsruher Instituts für Technologie (KIT)",
  year    = "2024",
  doi = {10.5445/IR/1000171925}
}

@misc{video_TDDFT,
	author = {Neepa Maitra},
	title = {Time-Dependent Density Functional Theory},
	howpublished = {\url{https://www.youtube.com/watch?v=ItWonHYm-ls}},
	year = {2010},
	note = {[Accessed 31-05-2025]},
  url = {https://www.youtube.com/watch?v=ItWonHYm-ls}
}

@ARTICLE{Grisafi2018-df,
  title     = "Symmetry-adapted machine learning for tensorial properties of
               atomistic systems",
  author    = "Grisafi, Andrea and Wilkins, David M and Cs{\'a}nyi, G{\'a}bor
               and Ceriotti, Michele",
  journal   = "Phys. Rev. Lett.",
  publisher = "American Physical Society (APS)",
  volume    =  120,
  number    =  3,
  month     =  jan,
  year      =  2018,
  copyright = "https://link.aps.org/licenses/aps-default-license",
  doi = {10.1103/PhysRevLett.120.036002}
}

@article{POT,
  author  = {Rémi Flamary and Nicolas Courty and Alexandre Gramfort and Mokhtar Z. Alaya and AurÃ©lie Boisbunon and Stanislas Chambon and Laetitia Chapel and Adrien Corenflos and Kilian Fatras and Nemo Fournier and LÃ©o Gautheron and Nathalie T.H. Gayraud and Hicham Janati and Alain Rakotomamonjy and Ievgen Redko and Antoine Rolet and Antony Schutz and Vivien Seguy and Danica J. Sutherland and Romain Tavenard and Alexander Tong and Titouan Vayer},
  title   = {POT: Python Optimal Transport},
  journal = {Journal of Machine Learning Research},
  year    = {2021},
  volume  = {22},
  number  = {78},
  pages   = {1--8},
  url     = {http://jmlr.org/papers/v22/20-451.html}
}

@ARTICLE{Waterman_Tmatrix,
  author={Waterman, P.C.},
  journal={Proceedings of the IEEE}, 
  title={Matrix formulation of electromagnetic scattering}, 
  year={1965},
  volume={53},
  number={8},
  pages={805-812},
  doi={10.1109/PROC.1965.4058}}

@article{Benedikt_homogenization,
author = {Zerulla, Benedikt and Venkitakrishnan, Ramakrishna and Beutel, Dominik and Krstić, Marjan and Holzer, Christof and Rockstuhl, Carsten and Fernandez-Corbaton, Ivan},
title = {A T-Matrix Based Approach to Homogenize Artificial Materials},
journal = {Advanced Optical Materials},
volume = {11},
number = {3},
pages = {2201564},
doi = {https://doi.org/10.1002/adom.202201564},
url = {https://onlinelibrary.wiley.com/doi/abs/10.1002/adom.202201564},
eprint = {https://onlinelibrary.wiley.com/doi/pdf/10.1002/adom.202201564},
year = {2023}
}

@article{Benedikt_multiscale,
author = {Zerulla, Benedikt and Krstić, Marjan and Beutel, Dominik and Holzer, Christof and Wöll, Christof and Rockstuhl, Carsten and Fernandez-Corbaton, Ivan},
title = {A Multi-Scale Approach for Modeling the Optical Response of Molecular Materials Inside Cavities},
journal = {Advanced Materials},
volume = {34},
number = {21},
pages = {2200350},
doi = {https://doi.org/10.1002/adma.202200350},
url = {https://onlinelibrary.wiley.com/doi/abs/10.1002/adma.202200350},
eprint = {https://onlinelibrary.wiley.com/doi/pdf/10.1002/adma.202200350},
year = {2022}
}

@article{treams_Beutel,
title = {treams – a T-matrix-based scattering code for nanophotonics},
journal = {Computer Physics Communications},
volume = {297},
pages = {109076},
year = {2024},
issn = {0010-4655},
doi = {https://doi.org/10.1016/j.cpc.2023.109076},
url = {https://www.sciencedirect.com/science/article/pii/S0010465523004216},
author = {Dominik Beutel and Ivan Fernandez-Corbaton and Carsten Rockstuhl}}

@software{Sax_Github,
  author = {Floris Laporte},
  title = {{SAX}: S-Matrices with Autograd and XLA - a scatter parameter circuit simulator and optimizer for the frequency domain based on JAX},
  url = {https://flaport.github.io/sax/},
  version = {0.14.1},
  year = {2024},
}

@article{Redheffer_product_original,
 ISSN = {00959057, 19435274},
 URL = {http://www.jstor.org/stable/24900576},
 author = {RAYMOND REDHEFFER},
 journal = {Journal of Mathematics and Mechanics},
 number = {3},
 pages = {349--367},
 publisher = {Indiana University Mathematics Department},
 title = {Inequalities for a Matrix Riccati Equation},
 urldate = {2026-02-06},
 volume = {8},
 year = {1959}
}

@article{Redhaffer_star_product,
    author = {Miguel Alexandre and Ivan M. Santos and Rodrigo Martins and Manuel J. Mendes},
    title = {SCATMM: Easy-to-Use Graphical User Interface for Light Propagation in Arbitrary Multilayers.},
    journal = {Journal of Open Research Software},
    volume = {12:11},
    year = {2024},
    doi = {https://doi.org/10.5334/ jors.511}
}

@article{S_matrix_Cotter,
author = {N. P. K. Cotter and T. W. Preist and J. R. Sambles},
journal = {J. Opt. Soc. Am. A},
number = {5},
pages = {1097--1103},
publisher = {Optica Publishing Group},
title = {Scattering-matrix approach to multilayer diffraction},
volume = {12},
month = may,
year = {1995},
url = {https://opg.optica.org/josaa/abstract.cfm?URI=josaa-12-5-1097},
doi = {10.1364/JOSAA.12.001097},
}

@article{S_matrix_STEFANOU199849,
title = {Heterostructures of photonic crystals: frequency bands and transmission coefficients},
journal = {Computer Physics Communications},
volume = {113},
number = {1},
pages = {49-77},
year = {1998},
issn = {0010-4655},
doi = {https://doi.org/10.1016/S0010-4655(98)00060-5},
url = {https://www.sciencedirect.com/science/article/pii/S0010465598000605},
author = {N. Stefanou and V. Yannopapas and A. Modinos},
}

@article{T_matrix_polarizabilities_Ivan,
author = {Fernandez-Corbaton, Ivan and Beutel, Dominik and Rockstuhl, Carsten and Pausch, Ansgar and Klopper, Wim},
title = {Computation of Electromagnetic Properties of Molecular Ensembles},
journal = {ChemPhysChem},
volume = {21},
number = {9},
pages = {878-887},
doi = {https://doi.org/10.1002/cphc.202000072},
url = {https://chemistry-europe.onlinelibrary.wiley.com/doi/abs/10.1002/cphc.202000072},
eprint = {https://chemistry-europe.onlinelibrary.wiley.com/doi/pdf/10.1002/cphc.202000072},
year = {2020}
}

@Article{Fernandez-Corbaton:2020,
author = {Fernandez-Corbaton,Ivan and Beutel,Dominik and Rockstuhl,Carsten and Pausch,Ansgar and Klopper,Wim},
title = {Computation of electromagnetic properties of molecular ensembles},
journal = {ChemPhysChem},
volume = {21},
pages = {878-887},
year = {2020},
doi = {10.1002/cphc.202000072}
}

@article{doi:10.1137/16M1106122,
author = {Nakatsukasa, Yuji and S\`{e}te, Olivier and Trefethen, Lloyd N.},
title = {The AAA Algorithm for Rational Approximation},
journal = {SIAM Journal on Scientific Computing},
volume = {40},
number = {3},
pages = {A1494-A1522},
year = {2018},
doi = {10.1137/16M1106122}
}

@article{https://doi.org/10.1002/lpor.202400584,
author = {Betz, Fridtjof and Hammerschmidt, Martin and Zschiedrich, Lin and Burger, Sven and Binkowski, Felix},
title = {Efficient Rational Approximation of Optical Response Functions with the AAA Algorithm},
journal = {Laser \& Photonics Reviews},
volume = {18},
number = {11},
pages = {2400584},
doi = {https://doi.org/10.1002/lpor.202400584},
year = {2024}
}

@article{https://doi.org/10.1002/adts.202400989,
author = {Fischbach, Jan David and Betz, Fridtjof and Asadova, Nigar and Tassan, Pietro and Urbonas, Darius and Stöferle, Thilo and Mahrt, Rainer F. and Burger, Sven and Rockstuhl, Carsten and Binkowski, Felix and Sturges, Thomas Jebb},
title = {A Framework to Compute Resonances Arising from Multiple Scattering},
journal = {Advanced Theory and Simulations},
volume = {8},
number = {2},
pages = {2400989},
doi = {https://doi.org/10.1002/adts.202400989},
year = {2025}
}

@article{adamoReliableDensityFunctional1999,
  doi = {10.1063/1.478522},
  title = {Toward Reliable Density Functional Methods without Adjustable Parameters: {{The PBE0}} Model},
  shorttitle = {Toward Reliable Density Functional Methods without Adjustable Parameters},
  author = {Adamo, Carlo and Barone, Vincenzo},
  year = {1999},
  journal = {The Journal of Chemical Physics},
  volume = {110},
  number = {13},
  pages = {6158--6170},

}

@article{eichkornAuxiliaryBasisSets1995,
  doi = {10.1016/0009-2614(95)00838-U},
  title = {Auxiliary Basis Sets to Approximate {{Coulomb}} Potentials ({{Chem}}. {{Phys}}. {{Letters}} 240 (1995) 283-290)},
  author = {Eichkorn, K and Treutler, O and {\"O}hm, H and H{\"a}ser, M and Ahlrichs, R},
  year = {1995},
  journal = {Chemical Physics Letters},
  volume = {242},
  number = {6},
  pages = {652--660},

}

@article{eichkornAuxiliaryBasisSets1997,
  doi = {10.1007/s002140050244},
  title = {Auxiliary Basis Sets for Main Row Atoms and Transition Metals and Their Use to Approximate {{Coulomb}} Potentials},
  author = {Eichkorn, Karin and Weigend, Florian and Treutler, Oliver and Ahlrichs, Reinhart},
  year = {1997},
  journal = {Theor Chem Acta},
  volume = {97},
  number = {1},
  pages = {119--124},

}

@article{ernzerhofAssessmentPerdewBurke1999,
  doi = {10.1063/1.478401},
  title = {Assessment of the {{Perdew}}\textendash{{Burke}}\textendash{{Ernzerhof}} Exchange-Correlation Functional},
  author = {Ernzerhof, Matthias and Scuseria, Gustavo E.},
  year = {1999},
  journal = {The Journal of Chemical Physics},
  volume = {110},
  number = {11},
  pages = {5029--5036},

}

@article{Holzer2020senex,
    author = {Holzer, Christof},
    title = "{An improved seminumerical Coulomb and exchange algorithm for properties and excited states in modern density functional theory}",
    journal = {J. Chem. Phys.},
    volume = {153},
    number = {18},
    pages = {184115},
    year = {2020},
    doi = {10.1063/5.0022755}
}

@article{https://doi.org/10.1002/jcc.22901,
author = {Plessow, Philipp and Weigend, Florian},
title = {Seminumerical calculation of the Hartree–Fock exchange matrix: Application to two-component procedures and efficient evaluation of local hybrid density functionals},
journal = {Journal of Computational Chemistry},
volume = {33},
number = {7},
pages = {810-816},
doi = {https://doi.org/10.1002/jcc.22901},
year = {2012}
}

@article{marij,
author = {Sierka,Marek  and Hogekamp,Annika  and Ahlrichs,Reinhart },
title = {Fast evaluation of the Coulomb potential for electron densities using multipole accelerated resolution of identity approximation},
journal = {J. Chem. Phys.},
volume = {118},
pages = {9136-9148},
year = {2003},
doi = {10.1063/1.1567253}

}

@article{PBE_a,
  title = {Generalized Gradient Approximation Made Simple},
  author = {Perdew, John P. and Burke, Kieron and Ernzerhof, Matthias},
  journal = {Phys. Rev. Lett.},
  volume = {77},
  pages = {3865--3868},
  year = {1996},
  doi = {10.1103/PhysRevLett.77.3865}
}

@article{PBE_b,
  title = {Generalized Gradient Approximation Made Simple [Phys. Rev. Lett. 77, 3865 (1996)]},
  author = {Perdew, John P. and Burke, Kieron and Ernzerhof, Matthias},
  journal = {Phys. Rev. Lett.},
  volume = {78},
  pages = {1396--1396},
  year = {1997},
  doi = {10.1103/PhysRevLett.78.1396}
}

@article{PM6_experimental_data_2023,
	author = {Piralaee, Mina and Asgari, Asghar},
	date = {2023/06/30},
	doi = {10.1007/s11082-023-05090-3},
	journal = {Optical and Quantum Electronics},
	number = {9},
	pages = {802},
	title = {The role of domain size and weight ratio of fullerene and non-fullerene acceptors on performance of PM6:Y6: PCBM ternary solar cell},
	url = {https://doi.org/10.1007/s11082-023-05090-3},
	volume = {55},
	year = {2023}
}

@article{ri,
author = {Ahlrichs, Reinhart},
title = {Efficient evaluation of three-center two-electron integrals over Gaussian functions},
journal  = {Phys. Chem. Chem. Phys.},
year = {2004},
volume = {6},
pages = {5119-5121},
doi = {10.1039/B413539C}
}

@article{TM_today_tomorrow,
  doi = {10.1021/acs.jctc.3c00347},
author = {Franzke, Yannick J. and Holzer, Christof and Andersen, Josefine H. and Begušić, Tomislav and Bruder, Florian and Coriani, Sonia and Della Sala, Fabio and Fabiano, Eduardo and Fedotov, Daniil A. and Fürst, Susanne and Gillhuber, Sebastian and Grotjahn, Robin and Kaupp, Martin and Kehry, Max and Krstić, Marjan and Mack, Fabian and Majumdar, Sourav and Nguyen, Brian D. and Parker, Shane M. and Pauly, Fabian and Pausch, Ansgar and Perlt, Eva and Phun, Gabriel S. and Rajabi, Ahmadreza and Rappoport, Dmitrij and Samal, Bibek and Schrader, Tim and Sharma, Manas and Tapavicza, Enrico and Treß, Robert S. and Voora, Vamsee and Wodyński, Artur and Yu, Jason M. and Zerulla, Benedikt and Furche, Filipp and Hättig, Christof and Sierka, Marek and Tew, David P. and Weigend, Florian},
title = {TURBOMOLE: Today and Tomorrow},
journal = {Journal of Chemical Theory and Computation},
volume = {19},
number = {20},
pages = {6859-6890},
year = {2023}

}

@misc{TURBOMOLE2023,
  title = {{TURBOMOLE V7.8 2023}, a development of {University of Karlsruhe} and {Forschungszentrum Karlsruhe GmbH}, 1989-2007, {TURBOMOLE GmbH}, since 2007},
  url = {https://www.turbomole.org},
  year = {2023}
}

@article{weigend2005a,
    author = {Weigend, Florian and Ahlrichs, Reinhart},
    title = {Balanced basis sets of split valence, triple zeta valence and quadruple zeta valence quality for H to Rn: Design and assessment of accuracy},
    journal = {Phys. Chem. Chem. Phys.},
    volume = {7},
    pages = {3297},
    year = {2005},
    doi = {10.1039/B508541A}
    
}

@article{weigendAccurateCoulombfittingBasis2006,
  doi = {10.1039/B515623H},
  title = {Accurate {{Coulomb-fitting}} Basis Sets for {{H}} to {{Rn}}},
  author = {Weigend, Florian},
  year = {2006},
  journal = {Phys. Chem. Chem. Phys.},
  volume = {8},
  number = {9},
  pages = {1057--1065},
 
}

@article{weigendBalancedBasisSets2005,
  doi = {10.1039/B508541A},
  title = {Balanced Basis Sets of Split Valence, Triple Zeta Valence and Quadruple Zeta Valence Quality for {{H}} to {{Rn}}: {{Design}} and Assessment of Accuracy},
  shorttitle = {Balanced Basis Sets of Split Valence, Triple Zeta Valence and Quadruple Zeta Valence Quality for {{H}} to {{Rn}}},
  author = {Weigend, Florian and Ahlrichs, Reinhart},
  year = {2005},
  journal = {Phys. Chem. Chem. Phys.},
  volume = {7},
  number = {18},
  pages = {3297--3305},

}

@article{YANAI200451,
title = {A new hybrid exchange–correlation functional using the Coulomb-attenuating method (CAM-B3LYP)},
journal = {Chemical Physics Letters},
volume = {393},
number = {1},
pages = {51-57},
year = {2004},
issn = {0009-2614},
doi = {https://doi.org/10.1016/j.cplett.2004.06.011},
author = {Takeshi Yanai and David P Tew and Nicholas C Handy}
}

@thesis{Kraus2025_Master,
  author  = {Kraus, Maria},
  title   = {Predicting Dynamic Polarizability Tensors of Organic Molecules Using Equivariant Graph Neural Networks},
  type    = {Master's thesis},
  institution = {Karlsruhe Institute of Technology (KIT)},
  year    = {2025}
}

@article{Hsu2026_TensorialSpectra,
  author  = {Hsu, Ting-Wei and Fang, Zhenyao and Bansil, Arun and Yan, Qimin},
  title   = {Accurate prediction of tensorial spectra using equivariant graph neural network},
  journal = {Nature Communications},
  year    = {2026},
  doi     = {10.1038/s41467-026-69159-9},
  volume = {17},
  pages = {3330}
}

@article{Nguyen2022_TensorialML,
  author  = {Nguyen, Vu Ha Anh and Lunghi, Alessandro},
  title   = {Predicting tensorial molecular properties with equivariant machine learning models},
  journal = {Physical Review B},
  volume  = {105},
  pages   = {165131},
  year    = {2022},
  doi     = {10.1103/PhysRevB.105.165131}
}

@misc{Gonnheimer2026_MACEMDP,
  author = {G{\"o}nnheimer, Nils and Reuter, Karsten and Kapil, Venkat and Margraf, Johannes T.},
  title  = {{MACE-MDP}: A General Dipole and Polarizability Model for Organic Molecules and Materials},
  year   = {2026},
  note   = {ChemRxiv preprint},
  doi    = {10.26434/chemrxiv.15000716/v2}
}

@article{Yang2019_StaticPol,
  author  = {Yang, Yang and Lao, Ka Un and Wilkins, David M. and Grisafi, Andrea and Ceriotti, Michele and DiStasio, Robert A.},
  title   = {Quantum mechanical static dipole polarizabilities in the {QM7b} and {AlphaML} showcase databases},
  journal = {Scientific Data},
  year    = {2019},
  note    = {PubMed ID: 31427579},
  volume = {6},
  number = {1},
  pages = {152}
}

@article{Wilkins2019_AlphaML,
  author  = {Wilkins, David M. and Grisafi, Andrea and Yang, Yang and Lao, Ka Un and DiStasio, Robert A. and Ceriotti, Michele},
  title   = {Accurate molecular polarizabilities with coupled cluster theory and machine learning},
  journal = {Proceedings of the National Academy of Sciences},
  year    = {2019},
  doi     = {10.1073/pnas.1816132116},
  volume = {116},
  number = {9},
  pages = {3401--3406}
}
